\documentclass{aa}
\usepackage{natbib}
\usepackage[version=4]{mhchem}
\usepackage[normalem]{ulem}
\usepackage{longtable}
\usepackage[dvipsnames]{xcolor}
\usepackage{caption}
\usepackage{placeins}
\usepackage{float}
\usepackage{textcomp}
\bibpunct{(}{)}{;}{a}{}{,} 
\usepackage{xcolor}
\usepackage{graphicx}
\usepackage{txfonts}
\usepackage{amsmath}    
\usepackage{amssymb}    
\begin{document} 

   \title{Observability of silicates in volatile atmospheres of super-Earths and sub-Neptunes}
    \subtitle{Exploring the edge of the evaporation desert}

   \author{M. Zilinskas
          \inst{1}
          \and
          Y. Miguel\inst{1,2} \and
          C.P.A. van Buchem\inst{1} \and
          I. A. G. Snellen\inst{1}.
          }

   \institute{Leiden Observatory, Leiden University, Niels Bohrweg 2, 2333CA Leiden, the Netherlands
   \and
                SRON Netherlands Institute for Space Research , Niels Bohrweg 4, 2333 CA Leiden, the Netherlands
                \\
            \\
              \email{zilinskas@strw.leidenuniv.nl}
             }

   \date{Received June XX, 2021; accepted July XX, 2021}

 
  \abstract
  {Many of the confirmed short-period super-Earths and smaller sub-Neptunes are sufficiently irradiated for the surface silicates to be sustained in a long-lasting molten state. While there is no direct evidence of magma ocean influence on exoplanets, theory suggests that, due to outgassing and diverse evolution paths, a wide range of resulting atmospheric compositions should be possible. Atmospheric contamination caused by the outgassing of the underlying magma ocean is potentially detectable using low-resolution spectroscopy. The James Webb Space Telescope provides the necessary spectral coverage and sensitivity to characterise smaller planets, including lava worlds. In light of this, we assess the observability of outgassed silicates submerged in volatile atmospheres on the edge of the evaporation valley. By placing a hypothetical 2 R${_\oplus}$ planet around a Sun-like star, we self-consistently model in 1D a wide range of potential atmospheric compositions, including thermal structure and outgassing. We focus on atmospheres rich in \ce{H}, \ce{C,} and \ce{N}. We assess the diverse chemistry of silicates and volatiles, and what features of outgassed species could be detected via emission spectroscopy using MIRI LRS. Results indicate that even for substantial volatile envelopes, strong in infrared opacity, the presence of silicates causes deep thermal inversions that affect emission. Similar to pure lava worlds, \ce{SiO} remains the only outgassed species with major infrared bands at 5 and 9 \textmu m. However, even a small amount of volatiles, especially of \ce{H2O} and \ce{H-}, may hinder its observability. We also find that the C/O ratio plays a large role in determining the abundance of \ce{SiO}. Detecting \ce{SiO} on a strongly irradiated planet could indicate an atmosphere with high metallicity and a low C/O ratio, which may be a result of efficient interaction between the atmosphere and the underlying melt.
  
  }
  


   \maketitle

\section{Introduction}

\begin{figure}
    \centering
        \includegraphics[width=0.49\textwidth]{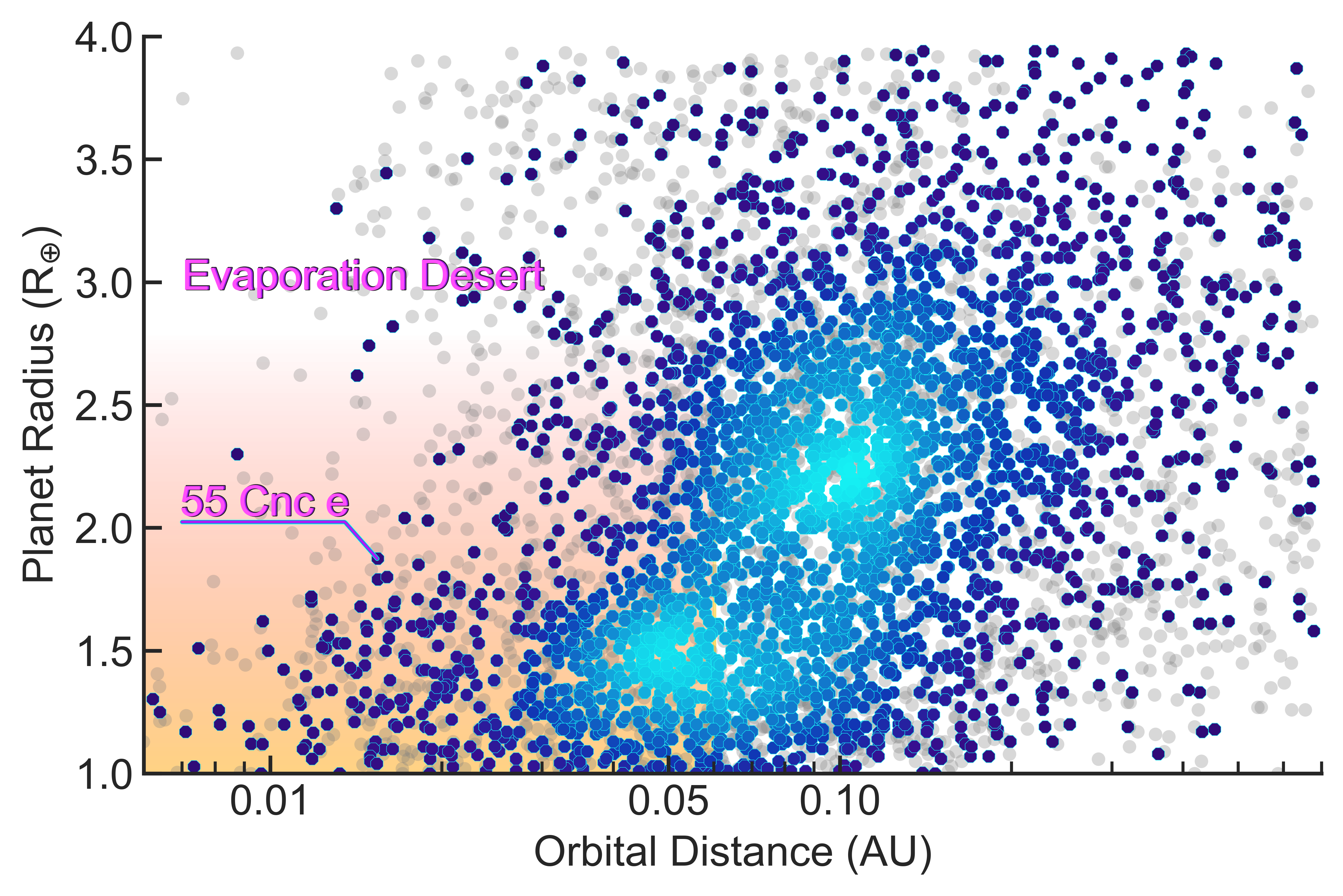}
    \caption{Short-period exoplanets with radii < 4 R${_\oplus}$. Coloured symbols indicate confirmed planets; grey symbols are candidate planets from the  Kepler, K2, and TESS missions. The colour value of confirmed planets represents the density of the occurrence rate, which peaks at two distinct radii (1.5 and 2.4 R${_\oplus}$), seemingly separating the population into super-Earths and sub-Neptunes. The highlighted region roughly encompasses the parameter space applicable to our modelled cases. Shown are the evaporation desert and one of the most well-studied super-Earths, 55 Cnc e. The data is  from the NASA exoplanet archive.}
    \label{fig:F1}
\end{figure}

Ever since their discovery, there has been great interest in trying to characterise and unravel the mysteries of the seemingly ambiguous super-Earths and sub-Neptunes. Although more massive Neptune-like planets are expected to retain most of the primordial H/He, intermediate (1.5-2.5 R${_\oplus}$) and smaller worlds are likely to be extremely diverse in their atmospheric composition and structure. Figure \ref{fig:F1} showcases the population of close-in planets.

Rocky worlds occupying the edge of the evaporation desert are shaped by erosion, accretion, degassing, and volcanism;  some of them possibly form long-lasting secondary atmospheres \citep{Elkins_2008}. Because of such close proximity to the star, many of the planets likely end up as bare rocks with no visible atmospheric component. Observations of several temperate and hot super-Earths seem to favour this theory \citep{Kreidberg_2019,Zieba_2022,Crossfield_2022}. However, even without an insulating atmosphere, they have temperatures high enough to engulf the dayside of the planet with magma oceans, which should result in tenuous but observable silicate envelopes \citep{Schaefer_2009,Miguel_2011,Ito_2015,Kite_2016,Zilinskas_2022}. For such worlds, \ce{SiO} and \ce{SiO2} have been proposed as the primary species that could be probed via infrared emission \citep{Ito_2015,Nguyen_2021,Zilinskas_2022}. It is also feasible that high-mean-molecular-weight species can survive erosion, leaving denser \ce{CO} or \ce{N2} atmospheres intact \citep{Zilinskas_2020}. A super-Earth 55 Cnc e may be a primary example of this \citep{Demory_2016b, Angelo_2017, Hammond_2017}. While studies indicate that planets below $\lesssim$ 2.0 R${_\oplus}$ are likely stripped of \ce{H2} \citep{Rogers_2021}, new interior models show that magma--atmosphere interaction during evolution could lead to large reservoirs of \ce{H2O}, buffering \ce{H2O} atmospheres, which, due to thermal and photochemical dissociation, should result in abundant \ce{H2} as a by-product \citep{Kite_2021,Dorn_2021}.  The only outstanding weakness of the proposed theory is the efficiency of the interaction between the melt and the atmosphere.

Going to larger radii (above 2.0 R${_\oplus}$), the observed discrepancy of densities may indicate the existence of water--ocean planets that are shrouded in dense steam atmospheres \citep{Zeng_2019,Mousis_2020,Nixon_2021,Bean_2021}. Insulation on sub-Neptunes is also expected to allow for deep magma oceans to be sustained indefinitely \citep{Kite_2020}. Just as with smaller planets, depending on the efficiency of the magma--vapour interaction and atmospheric mixing, it could result in \ce{H2}- or \ce{H2O}-rich envelopes that are heavily contaminated by silicate species \citep{Schlichting_2022,Kite_2020}.

Observations with JWST will provide the necessary constraints for the ongoing theoretical work. Extended \ce{H2} atmospheres of larger super-Earths and sub-Neptunes with substantial scale heights are easily probed via transmission spectroscopy \citep{Hu_2021_paper}. For intermediate and smaller planets, measuring emission of the dayside may prove to be the only viable characterisation method. However, even with JWST, characterising the chemistry of potential atmospheres will be challenging, detecting silicates even more so. From an observer's standpoint, finding whether these planets have atmospheres at all is a major stepping stone in the field of exoplanets. 

In this work we explore the chemistry and observability of outgassed silicates in volatile envelopes of irradiated rocky worlds. The highlighted region in Figure \ref{fig:F1} roughly indicates the parameter space applicable to this work. In contrast to studies done by \citet{Kite_2020}, \citet{Kite_2021}, and \citet{Schlichting_2022}, we do not model substantial atmospheres, but focus on cases where the surface pressure is relatively low in comparison to Neptune-like planets (< 10 bar). We make use of consistent outgassing equilibrium and radiative-transfer models to predict what silicate features are potentially characterisable through infrared emission spectroscopy, especially at wavelengths relevant for the  JWST MIRI instrument. Finding signs of silicates  could hint at an underlying magma ocean, allowing us to put better constraints on the proposed diversity of super-Earths and sub-Neptunes.

The paper is structured as follows. Section \ref{sec:Methods} contains the description of our approach in constructing  1D self-consistent atmospheric models, including chemistry and thermal structure. The analysis of the results is given in Section \ref{sec:Results}. We discuss some of the important factors that may affect observability in Section \ref{sec:Discussion}, and  conclude in Section \ref{sec:Conclusions}.

\section{Methods}
\label{sec:Methods}
\subsection{Setting up the system}
\label{sec:methodsIntro}
To explore observability of silicates in volatile atmospheres, we set up a grid of models that  represent a typical super-Earth or a sub-Neptune orbiting a Sun-like star. We focused on intermediate pressure envelopes in the range  1--10 bar. Our models focus on cases where the surface temperature is higher than 1400 K, which is enough for magma oceans to form and to influence the atmospheric composition. For a G-type star and dayside confined heat redistribution this typically results in a maximum orbital distance of 0.06 AU. We make use of several different codes that are set up to consistently calculate outgassing, chemical abundances, and temperature profiles, inclusive of the surface temperature. The models are then used to simulate emission spectra and expected JWST noise levels. Below we describe the approach for each of the components in more detail.

\subsection{Determining the chemistry}
\label{sec:methodsChem}
A major assumption made in this work is that the overlying atmosphere equilibrates with the molten surface, allowing outgassing to control the abundances of all silicates, including oxygen. Since the general atmospheric compositions of super-Earths and sub-Neptunes are unknown, we took the liberty of  exploring a grid of possible outcomes. These were drastically varied in metallicity, C/O ratio, volatile content, atmospheric pressure, and even internal temperature. 

For volatiles we took the solar composition \citep{Lodders_2009} and adjusted its metallicity (M/H), where all of the elements except for \ce{H} and \ce{O} were linearly scaled. While we normally assume that oxygen abundance is determined via outgassing, to explore the differences between solar and outgassed atmospheres we also modelled cases where oxygen is set by the metallicity parameter. In addition, we varied the C/O ratio (via carbon adjustment) together with the abundance of H/He, which allowed us to carefully dictate the major molecular constituents in the atmosphere. This allowed us to explore cases where strong irradiation and large sinks of light elements    (e.g. photoevaporation, dissolution) may result in high-mean-molecular-weight envelopes, dominated by  \ce{CO}, \ce{CO2}, or even \ce{N2}.


The outgassing budget was determined by the  open-source code \texttt{LavAtmos}\footnote{https://github.com/cvbuchem/LavAtmos} \citep{Buchem_2022}, which calculates the melt--vapour equilibrium for a given surface temperature and melt composition. To accurately determine the activity of the oxides in the melt, \texttt{LavAtmos} makes use of the liquid-solidus code \texttt{MELTS} \citep{Ghiorso_1995}. The package solves for the oxides containing the following elements: \ce{Al}, \ce{Ca}, \ce{Fe}, \ce{K}, \ce{Mg}, \ce{Na}, \ce{Si}, \ce{Ti}. The resulting outgassed partial pressures were added to the volatile atmospheres, while keeping the total surface pressure constant. This is equivalent to reducing the relative abundances of volatiles. As in \citet{Zilinskas_2022}, we took the magma to have the same composition as bulk silicate Earth (BSE). It contains 45.97 \% \ce{SiO2}, 36.66 \% \ce{MgO}, 8.24 \% \ce{FeO}, 4.77 \% \ce{Al2O3}, 3.78 \% \ce{CaO}, 0.35 \% \ce{Na2O}, 0.18 \% \ce{TiO}, and 0.04 \% \ce{K2O} (wt\%). The surface temperature and the outgassing were consistently calculated using a radiative-transfer code, as explained in Section \ref{sec:MethodsT}. It is important to note that currently \texttt{LavAtmos} does not account for the possible deposition of volatiles into the magma. As shown in the work of \citet{Kite_2020}, this can have substantial consequences on the atmospheric composition. The detailed analysis of this is beyond the  scope of this paper and will be a focus of a future study.

With the elemental budget determined, the atmospheric chemistry was solved using the thermochemical equilibrium code \texttt{FastChem}\footnote{https://github.com/exoclime/FastChem} \citep{Stock_2018,Stock_2022}. We took into account over 200 relevant species, inclusive of neutral and ion chemistry. The thermal data used was compiled from the Burcat NASA thermodynamics database.\footnote{http://garfield.chem.elte.hu/Burcat/burcat.html} 

\subsection{Computing thermal profiles}
\label{sec:MethodsT}


The temperature structure was solved using the radiative-transfer code \texttt{HELIOS}\footnote{https://github.com/exoclime/HELIOS} \citep{Malik_2017,Malik_2019}. We allowed convective adjustment to take place using an adiabatic coefficient of $\kappa = 2/7$, applicable to diatomic atmospheres. The profiles were treated with a rocky surface boundary, whose implementation  is described in detail in \citet{Malik_2019b,Whittaker_2022}. All of our models were reiterated until convergence such that the attained surface temperature is in good agreement with the atmospheric chemistry. For the purposes of showing possible spectral features, the heat redistribution was confined to the dayside of the planet (f=2/3). In specific cases it was approximated using the longwave optical depth of the atmosphere, based on equations from \citet{Koll_2022}.


We used a total of 50 opacity sources, including all of the major volatile and silicate species. The entire list and descriptions of all the opacities used in this study are given in Table \ref{table:opacities} in Appendix \ref{appendixA}. All of the atomic opacities are obtained from the DACE\footnote{https://dace.unige.ch/} database;   the majority use the Vienna Atomic Line Database (VALD3) line lists \citep{Ryab_2015}. For molecular opacities we made use of both the DACE database and the opacity calculator \texttt{HELIOS-K}\footnote{https://github.com/exoclime/HELIOS-K} \citep{Grimm_2015,Grimm_2021}. Following \citet{Grimm_2021}, the opacities were approximated using a Voigt fitting profile, wing cutting length of 100 cm$^{-1}$, and, where line lists allowed, a temperature of up to 4000 K.

\begin{figure}
    \centering
    \includegraphics[width=0.49\textwidth]{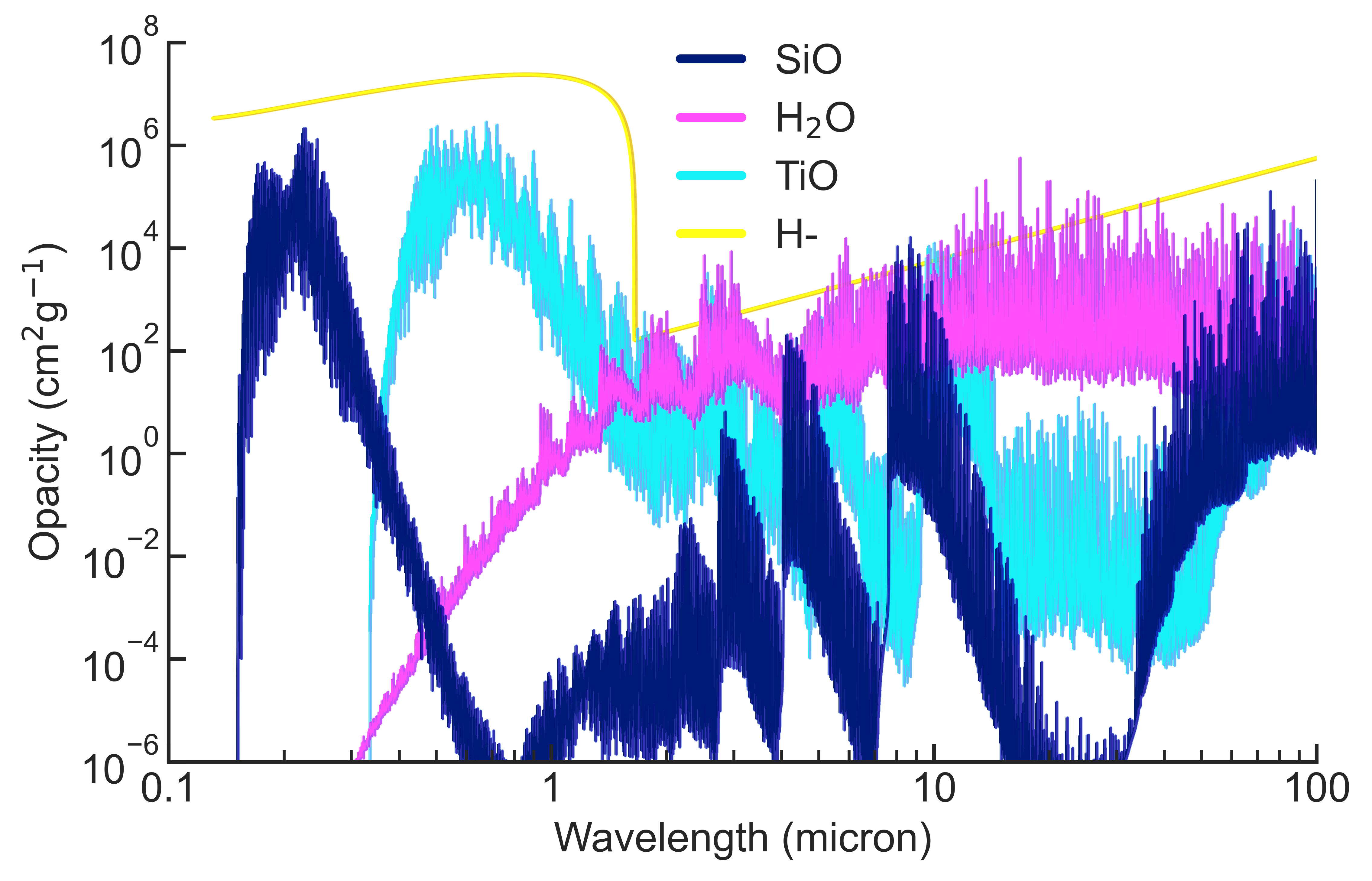}
    \caption{Comparison of \ce{SiO}, \ce{H2O}, \ce{TiO,} and \ce{H-} opacities, shown at a resolution of  $\lambda/\Delta\lambda = 2000$ for a temperature of 3000 K and atmospheric pressure of $10^{-2}$ bar. The description and sources of all  opacities used can be found in Table \ref{table:opacities}.}
\label{fig:F2}
\end{figure}

In terms of observability, \ce{SiO} is expected to be a key species of irradiated atmospheres \citep{Ito_2015,Zilinskas_2022}. In this work, we used the new ExoMol\footnote{https://www.exomol.com/} SiOUVenIR line list \citep{Yurchenko_2022}, which covers the entire UV--infrared wavelength range and is applicable to the high temperatures expected to occur on hot super-Earths. Figure \ref{fig:F2} shows the  key unweighted opacities considered in this work. \ce{SiO} shortwave opacity (< 1 \textmu m) is a strong contributor towards occurring temperature inversions, while the longwave bands peak at 5 and 10 \textmu m and are potentially detectable spectral features. While not displayed, there are many other potential species that are significant absorbers in silicate and volatile atmospheres.

For short-period planets shortwave stellar flux becomes an important factor in shaping the thermal structure of the atmosphere. Using simple black-body stellar models results in incorrect UV flux. Thus, all stellar irradiation models used in this work were generated via \texttt{HELIOS} using the PHOENIX \citep{Husser_2013} and MUSCLES \citep{France_2016,Youngblood_2016,Loyd_2016} databases. The spectra and opacities were sampled at a resolution of $\lambda/\Delta\lambda = 2000$ and cover the range of 0.1 -- 200 \textmu m.

\subsection{Simulating emission spectra}
\label{sec:MethodsE}

On hot super-Earths silicate atmospheres are expected to be confined to the tidally locked dayside of the planet, generally making them poor candidates for low-resolution transmission spectroscopy \citep{Zilinskas_2022}. Due to high atmospheric temperatures, spectral features may instead be probed through the emission of the secondary eclipse and phase curve measurements. If, however, these planets do possess global volatile atmospheres, transmission could be possible, but its viability will depend strongly on the scale height \citep{Zilinskas_2020}. While in this work we focus on emission spectroscopy, we note that for a number of known targets low-resolution transmission spectroscopy with JWST may also be feasible.

We generated emission spectra using the radiative-transfer code \texttt{petitRADTRANS}\footnote{http://gitlab.com/mauricemolli/petitRADTRANS} \citep{Molliere_2019,Molliere_2020}. We used the same atomic and molecular opacities described in Section \ref{sec:MethodsT}, including \ce{H2}, \ce{H2O}, and  \ce{O2} Rayleigh scattering, and \ce{H2-H2}, \ce{H2-He}, \ce{O2-O2}, and \ce{H-} continuum opacities. The spectra were calculated at a resolution of $\lambda/\Delta\lambda = 1000$ for a wavelength range of 0.3 -- 28 \textmu m, encompassing the coverage of all JWST instruments. In all figures the spectra were convolved to a lower resolution for better readability. 

For notable targets we assessed the  JWST noise using \texttt{PANDEXO}\footnote{https://exoctk.stsci.edu/pandexo/} \citep{Batalha_2017}, which is built on the Pandeia\footnote{http://jwst.etc.stsci.edu} engine. We only simulated MIRI Low Resolution Spectroscopy (MIRI LRS with $\lambda/\Delta\lambda \approx 100$) in slitless mode as it was likely to be the most suitable mode for the characterisation of silicate features. The wavelengths covered by the instrument are 5 -- 12 \textmu m. In each case we used the corresponding stellar and planetary parameters obtained from the NASA exoplanets archive. For the corresponding stellar models we used the   PHOENIX generated spectra.

\section{Results}
\label{sec:Results}

\subsection{Outgassed silicates in hydrogen atmospheres}
\label{sec:Results_I}

\begin{figure*}[ht]
    \centering
    \includegraphics[width=1\textwidth]{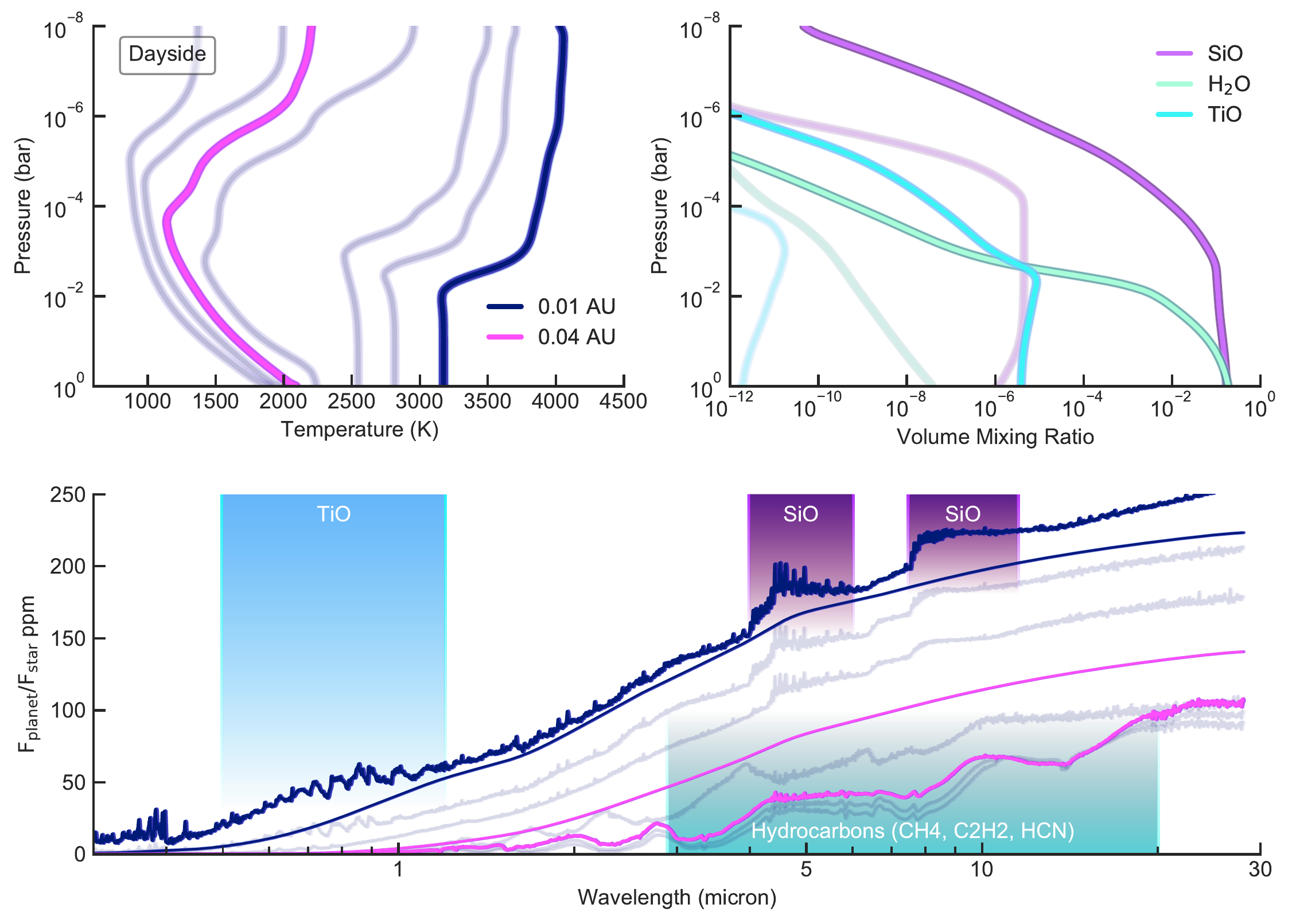}
    \caption{Atmospheric models of a super-Earth of 2 R${_\oplus}$ orbiting a Sun-like star. In all cases, dayside confined heat redistribution (f=2/3) and a surface pressure of 1 bar are assumed. The top left panel shows the temperature--pressure profiles at orbital distances of 0.01, 0.015, 0.02, 0.03, 0.04, 0.05, and 0.06 AU; the two highlighted cases are 0.01 AU (blue) and 0.04 AU (pink). In the top right panel the highlighted curves indicate abundances of \ce{SiO}, \ce{H2O}, and \ce{TiO} for the case of 0.01 AU with an effective planetary temperature of 3174 K. In the same panel the faded curves represent the chemistry of the same species at 0.04 AU (T${_{eff}}$ = 1771 K). The bottom panel contains the corresponding synthetic emission spectra, with the flat thinner curves representing black-body emission (assuming computed surface temperature). Major absorbers for the highlighted cases are indicated via shaded areas, with \ce{SiO}, \ce{TiO}, and hydrocarbons (\ce{CH4}, \ce{C2H2}, and \ce{HCN})   the primary species of interest. Spectra are shown at a resolution of $\lambda/\Delta\lambda = 600$.}
\label{fig:F3}
\end{figure*}

For a given initial composition the thermal structure and the resulting chemistry of an atmosphere is determined by the stellar flux that the planet receives. In Figure \ref{fig:F3} we showcase a hypothetical world placed around a Sun-like star of T = 5750 K. The only free parameter varied in each case is the orbital distance. The 2 R${_\oplus}$ planet is assumed to have a volatile-rich solar-like 1 bar atmosphere that is in equilibrium with an outgassed silicate component. In each model the  silicate abundance is computed via outgassing of a BSE melt of a numerically converged surface temperature. As expected, with increasing orbital distance the temperatures fall and the abundance of silicates decreases.

At close orbits the surface temperature can reach over 3000 K, which results in a substantial amount of outgassed \ce{O}, allowing for the  plentiful formation of oxides, including \ce{SiO} and \ce{H2O}. Our models show that 1 bar atmospheres with  surface temperatures higher than 2300--2500 K produce super-solar abundances of silicates, causing drastic changes in thermal structure. Shortwave absorbers heat the atmosphere causing deep thermal inversions, affecting even the photosphere. Below the photosphere, at around $10^{-2}$ bar, the atmosphere becomes optically thick to radiation, resulting in isothermal regions where no heat transport occurs (the three hottest temperature-pressure (TP) profiles in the top left panel of Fig. \ref{fig:F3}). A similar thermal structure is observed in volatile-free pure silicate atmospheres \citep{Zilinskas_2022}, implying that silicate opacities may largely be responsible in shaping the atmosphere.

Considering  the chemistry, we find that  many molecules survive thermal dissociation, even at the highest modelled temperatures. Abundances of major absorbers are shown in the top right panel of Fig. \ref{fig:F3}. While these atmospheres are filled with atomic species (\ce{H}, \ce{O}, \ce{Fe}, \ce{Mg,} and others), oxides, such as \ce{SiO}, \ce{H2O}, and \ce{TiO} dominate its opacity. At high pressures, \ce{H} and \ce{O} form \ce{H2O}, making it a strong absorber (see Fig. \ref{fig:F4}). Near the surface, \ce{H2O} and \ce{SiO} have similar volume mixing ratios. Moving to the upper low-pressure regions, \ce{H2O} begins to dissociate into atoms and ions, while \ce{SiO} remains in its molecular form, making it one of the most abundant species throughout the entire atmosphere. It should be expected that \ce{SiO} is a major constituent in atmospheres with an underlying magma ocean. For these cases we find that chemically the abundance of \ce{SiO} is only weakly affected by pre-existing volatiles. In addition to \ce{SiO}, the vaporisation of magma at high temperatures also results in high abundances of \ce{TiO}, which can become one of the most influential shortwave absorbers.

In the bottom panel of the figure we show the corresponding emission spectra. While in this case the small planet-to-star contrast results in a relatively low emission signal, the emergence of the 5 and 9 \textmu m \ce{SiO} features is clear (blue curve). Due to occurring inversions, \ce{SiO} increases the observed flux at these wavelengths. Unlike the silicate atmospheres modelled in \citet{Zilinskas_2022}, these show no significant sign of \ce{SiO2} absorption at 7 \textmu m. This is partly attributed to oxygen being chemically favoured to bond with volatiles such as hydrogen. At shorter wavelengths, \ce{TiO} is one of the dominant absorbers, causing a broad feature below 1 \textmu m. For BSE compositions, its presence is only important at high surface temperatures, typically greater than 2500 K. It is worth noting that many different molecules and atoms contribute to the shortwave opacity, some of which may be detectable in more extended atmospheres using transmission spectroscopy. Shortwave opacities are discussed in more detail at the end of this section. In addition to molecular opacities \ce{H-} becomes an important factor throughout the entire JWST wavelength range. Not only does it have a strong shortwave component, but its strong continuum at 10 \textmu m may also hinder observability of \ce{SiO}. Overall, out of all the outgassed silicates, the two \ce{SiO} features are likely to be the easiest to characterise using MIRI LRS covering the 5--12 \textmu m range.

Moving to colder cases, the abundance of all silicates decreases rapidly, becoming sub-solar at 0.04 AU (pink curve). The total outgassed pressure of just silicates at temperatures below 2500 K is comparable to 1 millibar \citep{Zilinskas_2022}. Assuming melt--vapour equilibrium is attained, the volatile component is likely to dominate, making species such as \ce{SiO} or \ce{TiO} unobservable with low-resolution spectroscopy. Another consequence of this is a drastic reduction of outgassed oxygen, which raises the C/O ratio, allowing hydrocarbons to efficiently form. Most of the species in cooler atmospheres are heavily weighted towards infrared opacity, resulting in a lack of any significant inversions that may impact observability. The spectrum here is dominated by molecules such as \ce{CH4}, \ce{C2H2}, and \ce{HCN}, all showing deep absorption features. Detecting silicates in emission at relatively large orbits could indicate that either the temperature of the melt is much higher than the planetary equilibrium temperature or that silicates are not in equilibrium with the underlying melt. 

\subsection{Contribution function}
\begin{figure*}
    \centering
    \includegraphics[width=1.0\textwidth]{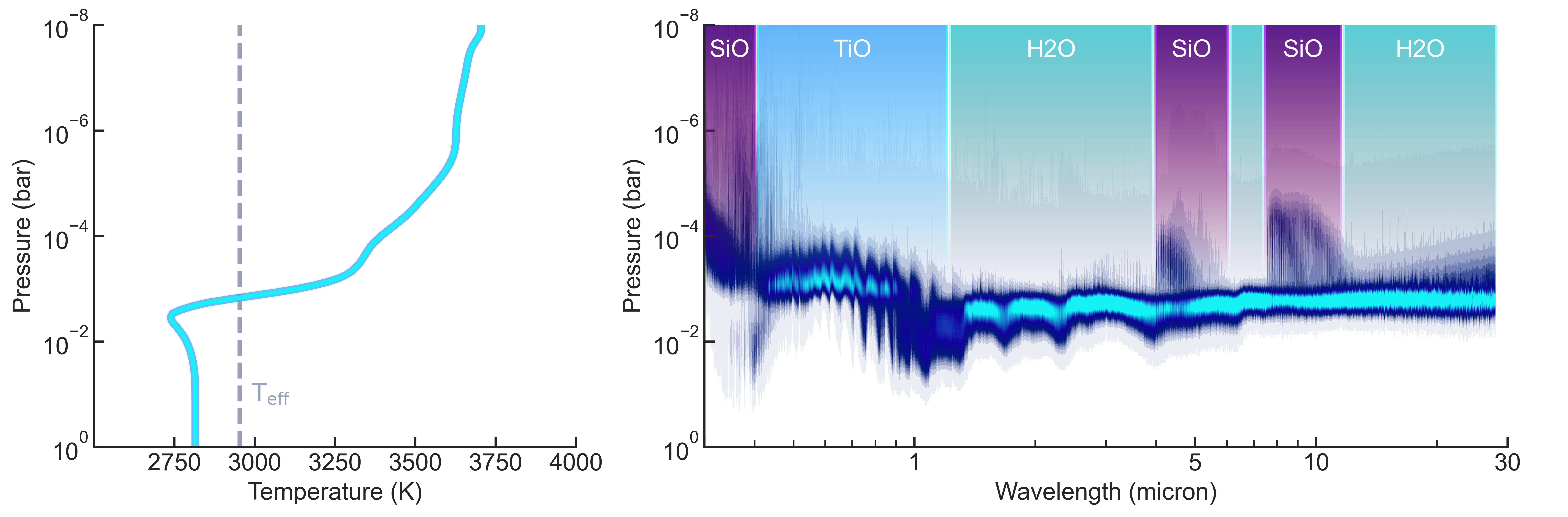}
    \caption{Emission contribution function of a strongly irradiated super-Earth orbiting a Sun-like star at 0.015 AU. The temperature profile in the left panel is   from the models showcased in Fig. \ref{fig:F3}. The dashed line is the effective temperature of the planet T = 2950 K. The right panel show the emitting region of the atmosphere as a function of wavelength. Major contributing molecules are indicated in their respective regions. Lesser contributing opacities are discussed in the text. Spectra are shown at a resolution of $\lambda/\Delta\lambda = 600$.}
\label{fig:F4}
\end{figure*}

In Figure \ref{fig:F4} we take one of the strongly irradiated cases and show its emission contribution function. In the right panel the highlighted region represents the emitting photosphere. For wavelengths > 1 \textmu m, this mostly coincides with pressures between $10^{-4}$ and $10^{-2}$ bar. A major contributing molecule for longwave opacity is \ce{H2O}. Its dominance is a general occurrence in our models. Plentiful hydrogen and oxygen ensure that even at high temperatures it is one of the most optically dominating species. The  leftover hydrogen results in a strong \ce{H-} continuum, pushing the general opacity higher. The tail of the continuum can be seen at wavelengths > 10 \textmu m. Since the abundance of \ce{SiO} is not strongly affected by increasing temperatures and lower pressures, its opacity has large contributions from inverted regions. If atmospheres of super-Earths are prone to thermal inversions, it is likely that \ce{SiO} will show up as increased flux. The 9 \textmu m feature is and should be visible, even with strong volatile opacities present. If no volatiles are present, enough \ce{SiO2} may form to appear at 7 \textmu m, complementing the \ce{SiO} feature \citep{Zilinskas_2022}.

There are many different species contributing to the total shortwave opacity (< 1 \textmu m).
\ce{SiO}, \ce{AlO}, \ce{MgO}, \ce{TiO}, \ce{Mg}, and \ce{Fe} all have very strong opacities. Some lesser, but important species are \ce{SiH}, \ce{MgH}, \ce{VO}, \ce{Al}, \ce{Ca}, \ce{K}, \ce{Na}, \ce{Si}, and \ce{Ti}. \ce{TiO}, which has a broad wavelength coverage, is perhaps the most important for observations, and for  its influence in shaping the thermal structure. Its presence is known to strongly affect atmospheres, even in gas giants \citep{Serindag_2021}. We note that on rocky planets, \ce{TiO} is typically sustained in significant abundances only above 2500--2800 K. Aside from \ce{TiO}, the \ce{SiO} UV band and \ce{Fe} opacity have major influence on the strength and depth of the occurring inversions. These species are also much more volatile and readily vaporised from the magma. It is important to note that because of the large number of shortwave absorbers, even atmospheres that are missing major oxides such as \ce{SiO} or \ce{TiO} can still have deep occurring inversions. 

Previous studies have shown that pure silicate atmospheres have similar total shortwave opacity \citep{Zilinskas_2022}. This is unsurprising since the majority of shortwave absorbers come from silicate outgassing. While there are additional shortwave absorbers due to the presence of volatiles, namely \ce{SiH}, \ce{MgH}, and \ce{VO}, these are relatively minor in comparison to silicates. We note that, due to a lack of thermal data, our models do not include \ce{FeH}, whose   opacity     peaks at 1 \textmu m. For atomic species we also do not use pressure-caused broadening, likely leading to some underestimation of the line widths. It is possible that many of the atomic species, especially alkali metals, are much more dominant in shaping the atmosphere. 

The inherent complexity of the shortwave region makes it difficult to correctly model temperature profiles. Many of the mentioned opacities here are often overlooked, leading to theoretically incorrect temperatures. After the chemistry, the shortwave opacities are likely to be a major source of uncertainties that can greatly affect the interpretations of observed spectra.

\subsection{Impact of metallicity and C/O ratio}

\begin{figure}
    \centering
    \includegraphics[width=0.49\textwidth]{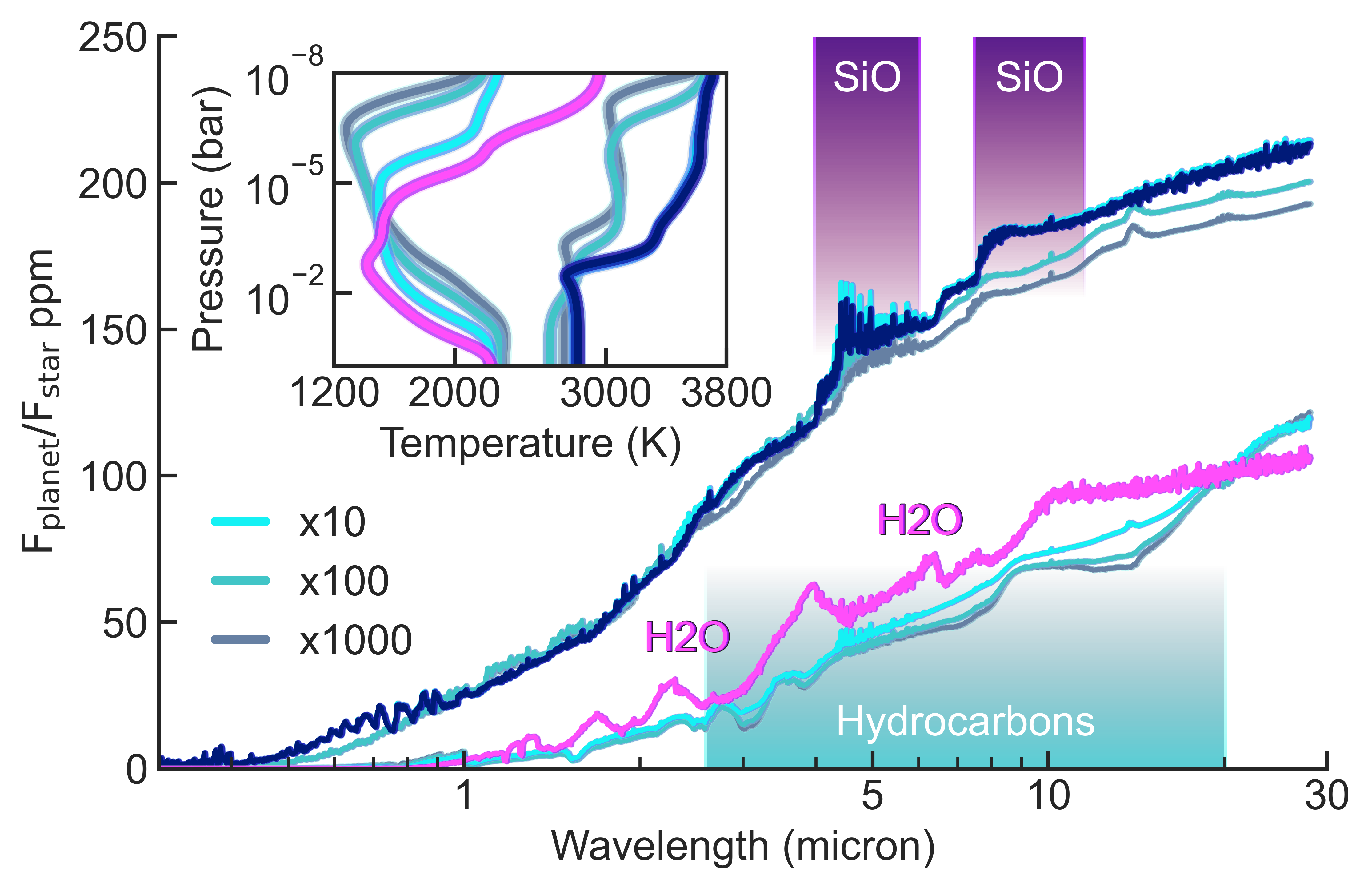}
    \caption{Synthetic emission spectra for an atmosphere of increased metallicity. The main cases (in  blue and pink) represent models with solar metallicity at two different orbital distances (0.015 and 0.03 AU). For each orbit, atmospheres of 10, 100, and 1000 times metallicity are shown. Some of the contributing opacities are shown for their respective wavelengths. The inset displays the corresponding temperature temperature profiles. The   metallicity here does not control the abundance of outgassed silicates or oxygen. Spectra are shown at a resolution of $\lambda/\Delta\lambda = 600$.}
\label{fig:F5}
\end{figure}

\begin{figure*}
    \centering
    \includegraphics[width=1.0\textwidth]{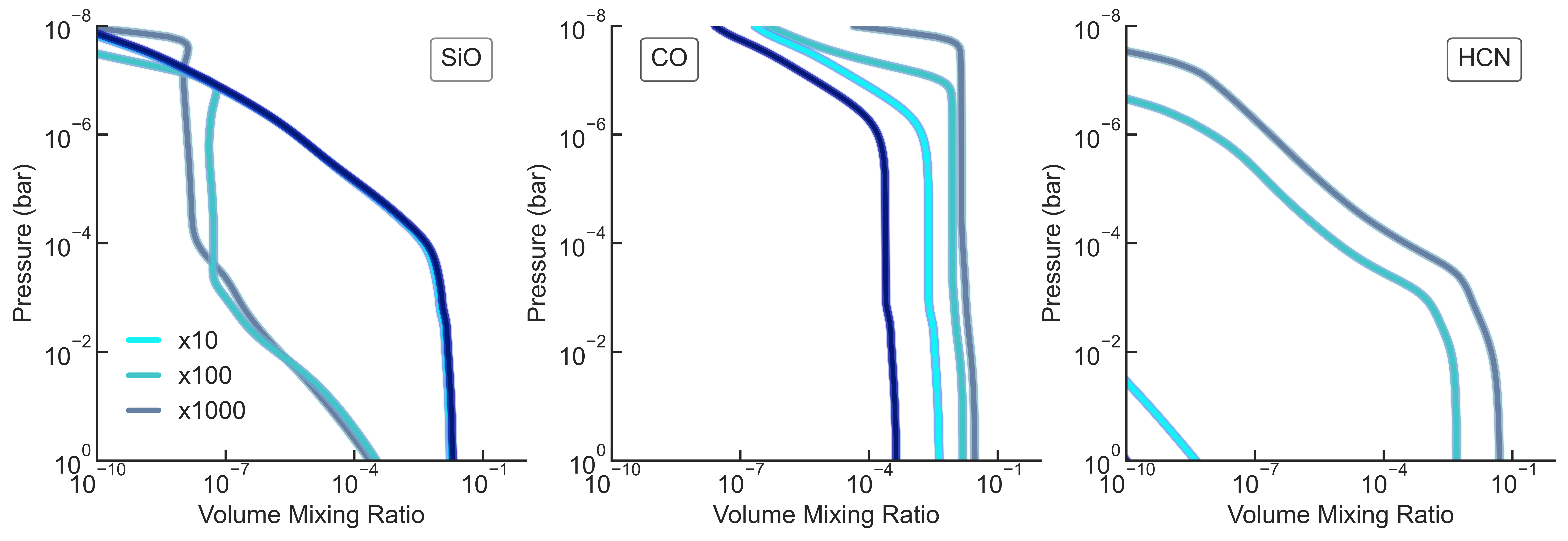}
    \caption{Volume mixing ratios of \ce{SiO}, \ce{CO}, and \ce{HCN} (from left to right). The models shown here are for the close orbit (0.015 AU) cases in Fig. \ref{fig:F5}. Each panel contains the original solar metallicity (blue) as well as 10, 100, and 1000 times increased metallicity curves.}
\label{fig:F5Chem}
\end{figure*}

The formation process  for short-period rocky planets is unknown, but it is often assumed that they are heavily enriched in metals \citep{Weiss_2013,Moses_2013}. In Figure \ref{fig:F5} we  vary the  metallicity to explore what effect it may have on the observability of silicates. The blue and pink curves represent the original solar models discussed in the previous section, while for each of the orbits the overplotted curves show atmospheres with 10, 100, and 1000 times increased metallicity. We note that the metallicity here does not control the abundances of outgassed silicates or oxygen, but only of the  volatiles. The main impact is thus the  increase in \ce{C} and  \ce{N, } and a higher C/O ratio. The corresponding chemistry of the close orbit cases is shown in Figure \ref{fig:F5Chem}.

\begin{figure}
    \centering
    \includegraphics[width=0.49\textwidth]{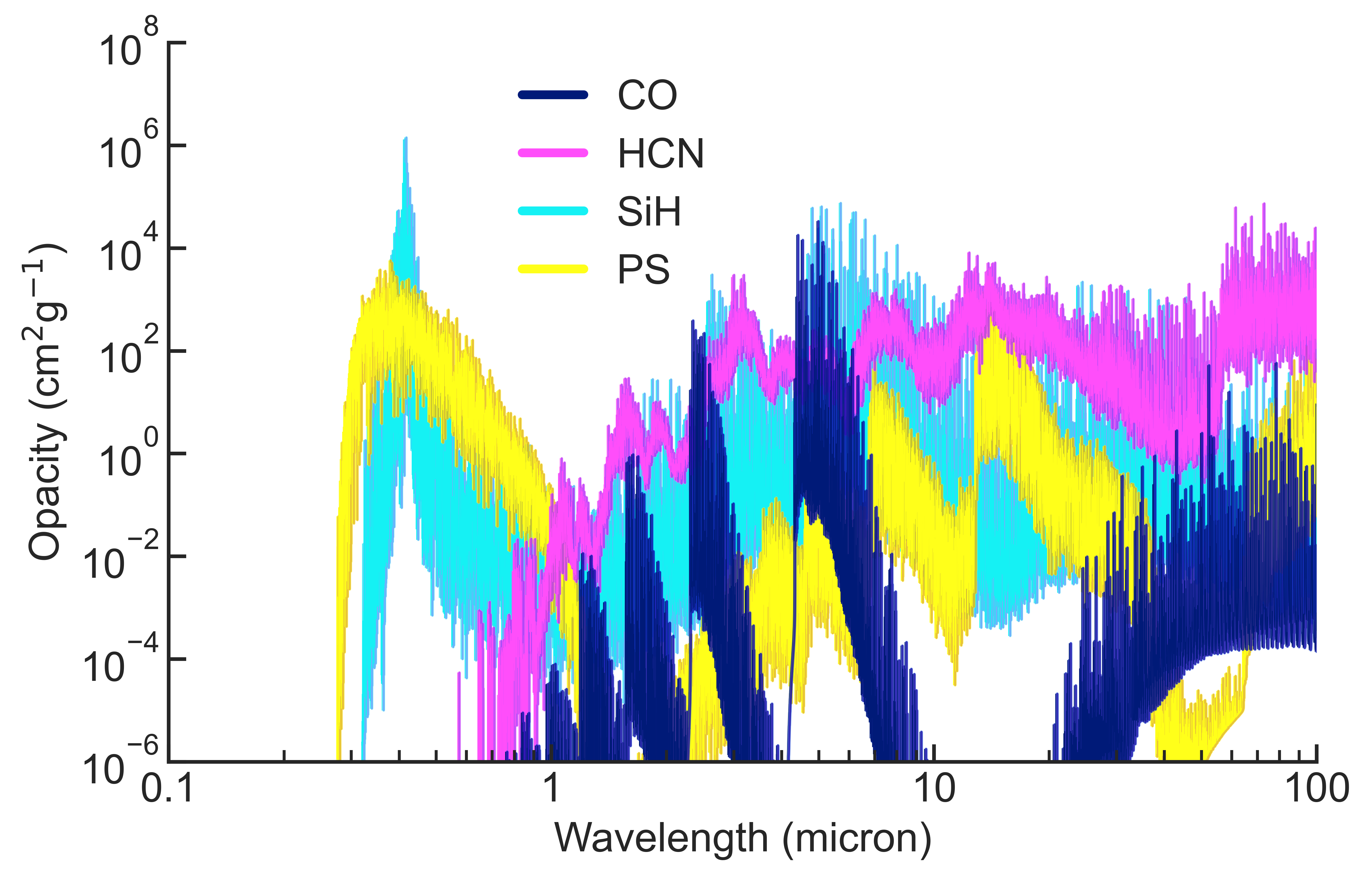}
    \caption{Unweighted abundance  opacities of \ce{CO}, \ce{HCN}, \ce{SiH,} and \ce{PS}, shown at a resolution of  $\lambda/\Delta\lambda = 2000$ for a temperature of 3000 K and atmospheric pressure of $10^{-2}$ bar. Detailed description of all opacities used in this work can be found in Table \ref{table:opacities} in Appendix \ref{appendixA}.}
\label{fig:F2Extra}
\end{figure}

For close orbits an increase of about  x10  in metallicity has a minimum effect on the atmospheric chemistry or thermal structure. With \ce{SiO} and \ce{H2O} remaining as dominant oxides, the spectral features are mostly unchanged. This slight increase in metallicity  allows   \ce{CO} to form more efficiently, very slightly boosting its opacity at 5 \textmu m. The unweighted opacity of \ce{CO}, along with a few other species that are discussed later, are shown in Figure \ref{fig:F2Extra}. When the metallicity is increased to x100, the C/O ratio crosses unity and the chemistry starts prioritising the formation of \ce{CO}, heavily diminishing other oxides (see Fig. \ref{fig:F5Chem}). Atmospheres that do not outgas or retain enough oxygen are likely to suffer this effect, erasing opacities of \ce{SiO} or \ce{H2O} in the spectrum. With no \ce{SiO}, \ce{Si} either remains in atomic form or bonds with \ce{H} to form \ce{SiH}. However, due to abundant \ce{N} and the high C/O ratio, \ce{H} prioritises bonding with \ce{CN} to form \ce{HCN} (rightmost panel of Fig. \ref{fig:F5Chem}). This chemistry is now reflected in the thermal structure as inversions become significantly weaker. Pushing the metallicity higher further increases the C/O ratio, resulting in a mostly \ce{CO} and hydrocarbon-dominated atmosphere, even at high temperatures. Because the emitting photosphere resides mostly in the isothermal region, the spectrum becomes largely featureless. 

With increasing orbital distance (pink curve in Fig. \ref{fig:F5}), the trends in the chemistry and spectra remain similar, but more severe. Since the abundance of oxygen from outgassing is low at these temperatures, the C/O ratio at x1 metallicity  is already near unity. Even at x10 the C/O ratio becomes much higher than unity causing efficient formation of hydrocarbons. The dominant molecules become \ce{HCN}, \ce{C2H2}, and \ce{CO}, while \ce{H2O} and any potential \ce{SiO} are erased from the atmosphere. This results in opacity values  that are  heavily weighted towards infrared wavelengths, and  thus a lack of deep inversions.

\subsection{Keeping the C/O ratio constant with metallicity}

\begin{figure}
    \centering
    \includegraphics[width=0.49\textwidth]{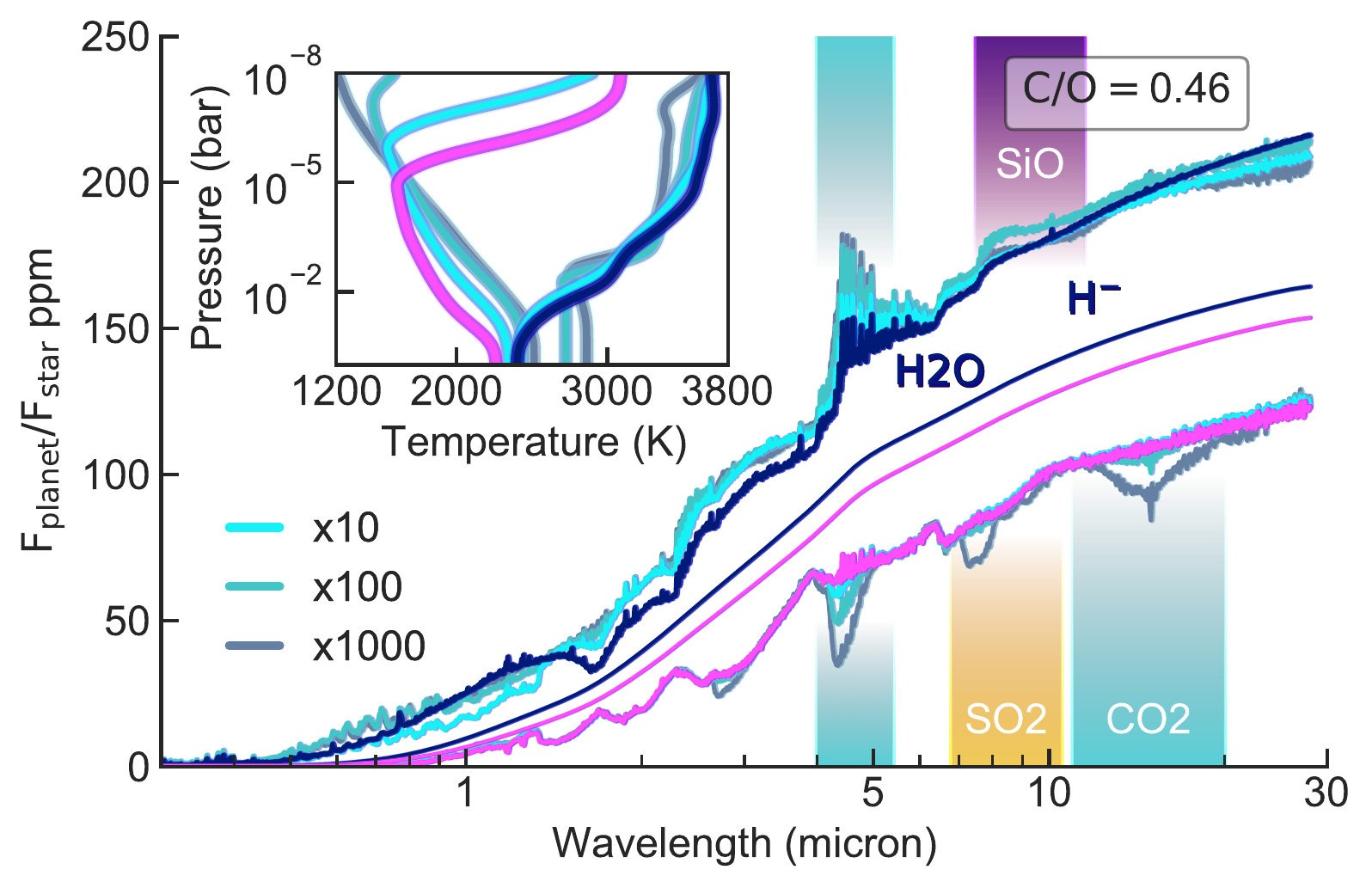}
    \caption{Synthetic emission spectra as in Fig. \ref{fig:F5}, but with constant C/O ratio of 0.46. The oxygen abundance here is scaled with metallicity. The   blue and and pink curves represent cases with solar metallicity. The respective black-body emission is represented as thin curves in the same colour. Major contributing opacities for solar cases are indicated with shaded regions. The inset shows the  corresponding temperature profiles. The surface temperature increases with metallicity, shifting temperature-pressure profiles to the right. Spectra are shown at a resolution of $\lambda/\Delta\lambda = 600$.}
\label{fig:F7}
\end{figure}

The balance between carbon and oxygen is a major factor in determining atmospheric chemistry and whether \ce{SiO} is allowed to thrive. While in Figures \ref{fig:F5} and \ref{fig:F5Chem} we allow outgassing to control the abundances of oxygen, and therefore the C/O ratio, in Figure \ref{fig:F7} we set a constant C/O ratio. The value of oxygen is now scaled with metallicity. The blue and pink curves represent models at the same orbital distances as in the previous figures with cases of 10, 100, and 1000 times increased metallicity shown for each.

The abundance of oxygen at solar value is an order of magnitude lower than what would be outgassed in our close orbit case (blue curve). This directly results in decreased abundances of all oxides, inclusive of \ce{SiO}. Additionally, the solar C/O ratio is high enough for \ce{CO} to form and become an abundant oxygen carrier in the atmosphere. Other oxides, such as \ce{CO2}, \ce{H2O}, or \ce{SiO} follow closely behind. The resulting infrared opacity is dominated by the continuum of \ce{H-}, with some contribution from \ce{H2O}. Overshadowed, but significant are opacities of \ce{SiO} and \ce{CO2}. The atmosphere in this case becomes less opaque, allowing radiation to penetrate deeper. This causes inversions to extend nearly all the way to the surface. To conserve energy, the surface temperature is consequently decreased. Increasing metallicity to 100x results in oxygen abundances similar to those produced by  a melt-vapour equilibrium. This is reflected in the drastic change in the temperature profile, which now resembles cases presented in the  previous figures (blue curve in Fig. \ref{fig:F5}). The high surface temperature also produces more \ce{Si}, which is why we see its features begin to emerge. Increasing metallicity further has a similar effect. Regardless of metallicity in these models, a solar C/O ratio makes \ce{CO} a prolific species, which absorbs much of the oxygen and reduces the possibility of \ce{SiO} being a dominant species.

With outgassing lessened at larger orbits (pink curve), \ce{H2O} and \ce{CO} are the main oxygen carriers. In contrast to previous cases, increasing metallicity here results in high abundances of \ce{CO2}, which has several important bands in the infrared,  the  two most important being at 4.5 and 14 \textmu m. With high metallicity \ce{SiO} reaches a volume mixing ratio of nearly $10^{-3}$, but the opacity of \ce{H2O} completely overshadows it, making it invisible. We additionally see the emergence of some lesser species, such as \ce{SO2}, which   has a strong opacity at 7.5 \textmu m.

In summary, our models indicate that it is mainly the C/O ratio that significantly affects atmospheric chemistry, including the final abundance of \ce{SiO}; the  metallicity of volatiles is much less important. Therefore, lava worlds enveloped in volatiles are likely to depend heavily on the balance between carbon and oxygen. High C/O ratios drive oxygen atoms away from \ce{SiO}, potentially making \ce{SiH} a species of interest. As shown in Fig. \ref{fig:F2Extra}, \ce{SiH} has a strong feature at 0.4 \textmu m and several other bands between 1 and 10 \textmu m. Nonetheless, in hydrogen-rich worlds, regardless of the C/O ratio, \ce{H2O} and \ce{H-} may further hinder observability of silicate species.


\subsection{Effect of increased surface pressure and internal heating}

\begin{figure}
    \centering
    \includegraphics[width=0.49\textwidth]{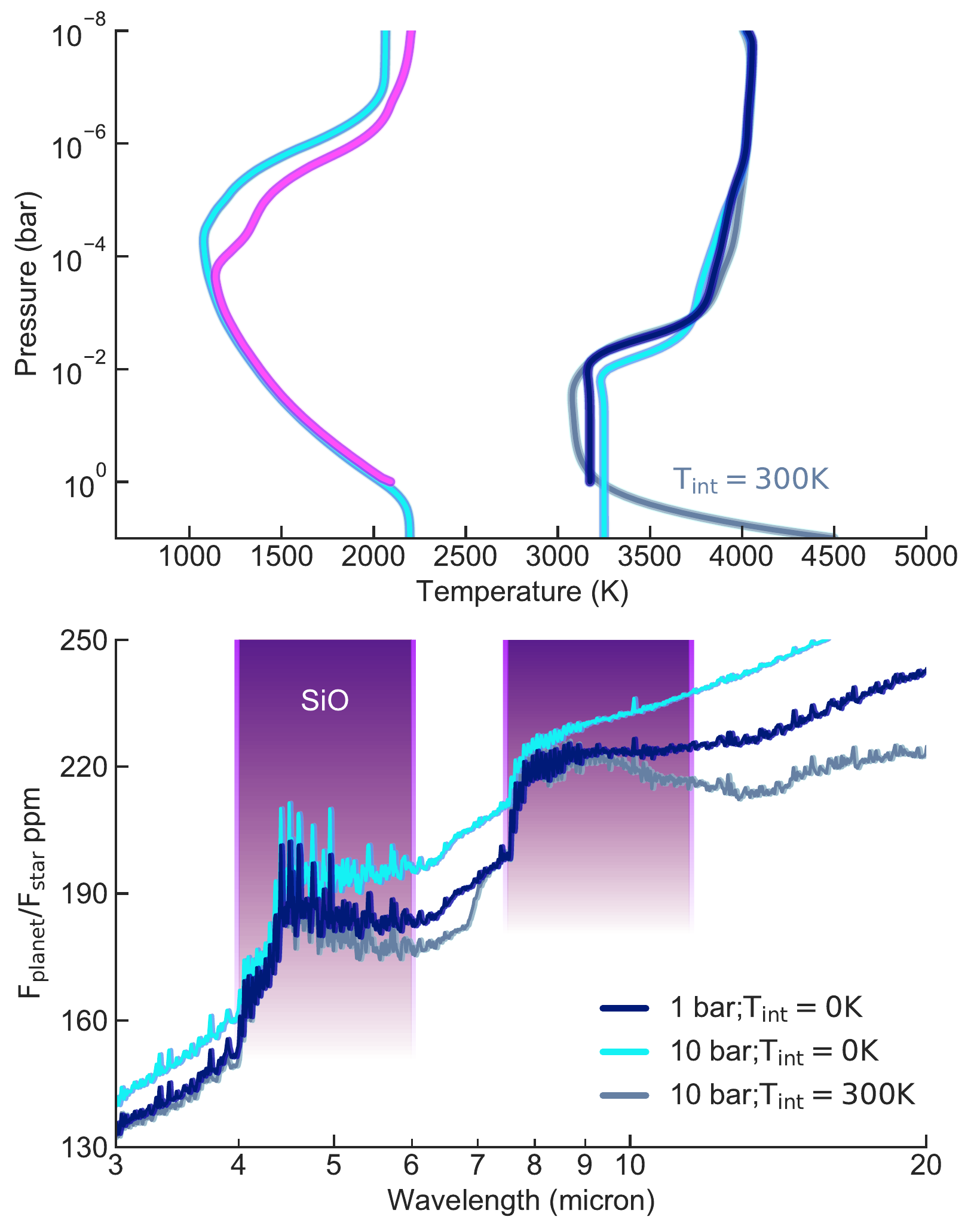}
    \caption{Temperature--pressure profiles and emission spectra of varied atmospheric pressure and internal heating. The top panel contains profiles of atmospheres with 1 and 10 bar at two different orbital distances (0.01 and 0.04 AU). For the close orbit, an additional case with T$_{int}$=300 K is shown. The lower panel shows the resulting emission spectrum of the close orbit cases. The chosen wavelength region is where \ce{SiO} features are expected to appear. Spectra are shown at a resolution of $\lambda/\Delta\lambda = 600$.}
\label{fig:F6}
\end{figure}

Down from millibar silicate atmospheres to volatile-shrouded sub-Neptunes, the envelopes of rocky worlds may vary by orders of magnitude in surface pressure. Larger thermally thick atmospheres can act as insulators, allowing the molten state of a surface to exist indefinitely, even if the planet is weakly irradiated \citep{Lopez_2014,Kite_2020}. One resulting consequence of this may be that insulation and trapped heat near the surface allow for much greater melt temperatures, making silicates more dominant over volatiles. It is also possible that due to an  increase in  volatiles, the outgassing becomes heavily suppressed. In the previous sections we  assume the volatiles to be capped at 1 bar, while also keeping the internal temperature of the planet at 0 K. While the exact dynamics of this are beyond the  scope of this paper, in Figure \ref{fig:F6} we show how an increase in pressure and internal heating might affect the observability of silicon-bearing species.

The bright cyan curves in the top panel of the figure represent atmospheres with a surface pressure increased to 10 bar. Without the internal temperature enabled, there is no additional heat to transport, making convection unnecessary, which results in a simple extension of the isothermal region. However, if the atmosphere is supplied with energy from below, the optically and thermally thick portion becomes unstable to convection, dramatically increasing the temperature of the melt. Because of the exponential scaling of outgassing, this can often lead to an immense increase in silicate abundances. If these atmospheres are well mixed, we should expect to see substantial silicate features.

The bottom panel of Figure \ref{fig:F6} shows the spectra for the close orbit models. A 10 bar atmosphere with no internal heating results in a greater dominance of the volatile component. The \ce{H-} continuum becomes more effective, concealing the presence of \ce{SiO}. With T$_{int}$=300 K, even at 10 bar of volatile content, \ce{SiO} becomes the most abundant species in the atmosphere, causing its features to dominate the spectrum. We note that our arbitrary use of unusually high T$_{int}$ is purely for demonstrative purposes. Close-in rocky planets are susceptible to various heating mechanisms outside stellar irradiation. With an insulating optically thick atmosphere, it is possible that the trapping of heat allows global magma oceans to be sustained at much hotter temperatures than expected. Observations of silicates can therefore provide important constraints on the properties of the melt and the interior \citep{Zilinskas_2022}.

\subsection{Depletion of hydrogen}

\begin{figure*}[!h]
    \centering
    \includegraphics[width=1.0\textwidth]{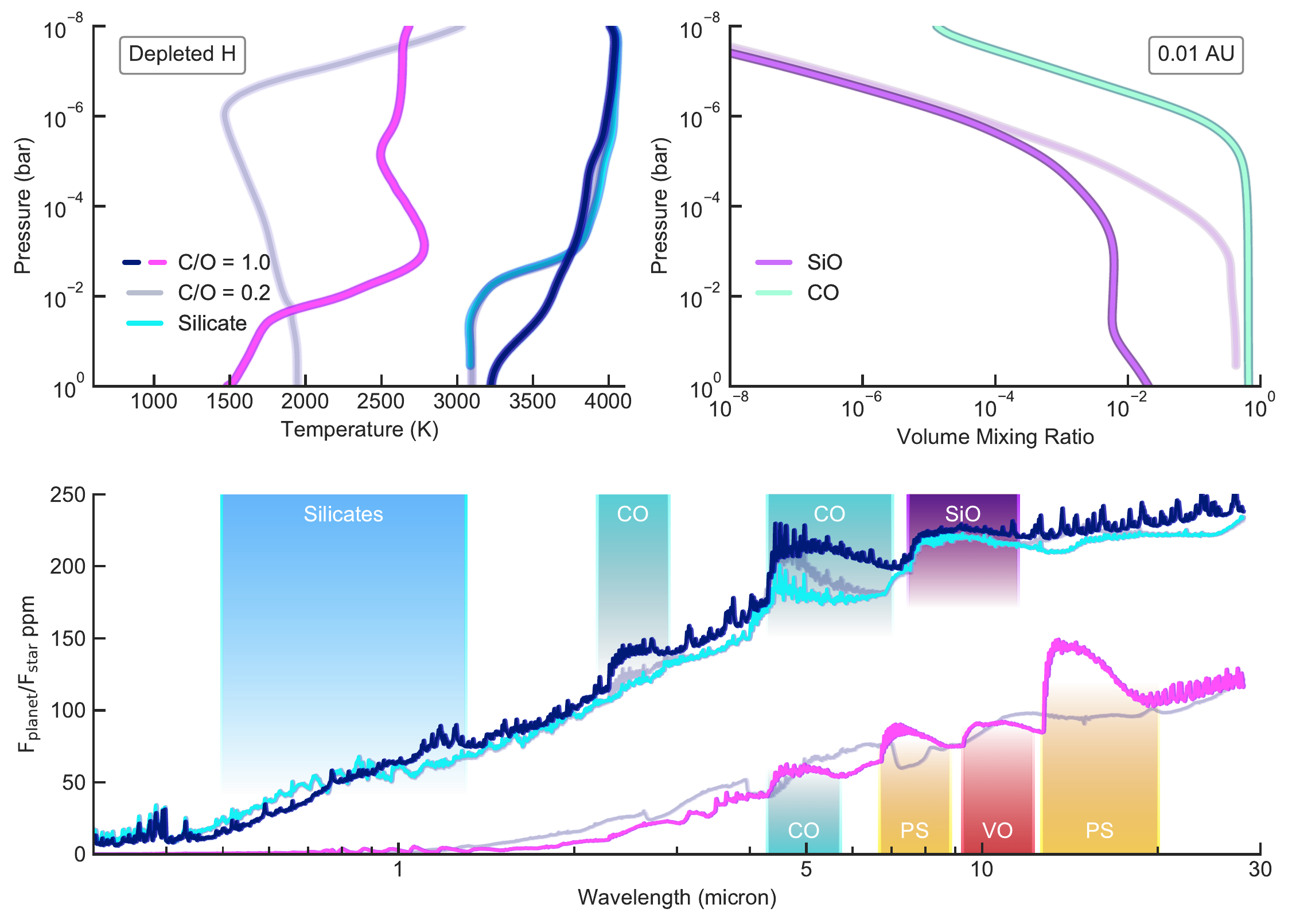}
    \caption{Atmospheric models with severely depleted hydrogen at orbital distances of 0.01 and 0.04 AU. Top panel:  TP profiles for C/O ratios of 1.0 (pink and blue) and 0.2 (faded) and an additional model of a pure silicate atmosphere (cyan).  All cases assume dayside heat redistribution (f=2/3) with volatile-containing atmospheres having 1 bar surface pressure. The total pressure in the pure silicate case is computed using the outgassing code. Top right:     Abundances of \ce{SiO} and \ce{CO} for the 0.01 AU case (blue TP profile), with the faint curves representing a pure silicate atmosphere. Bottom panel: Emission spectra colour-coded for their respective TP profiles. The labelled regions denote major opacity contributions for the cases with a C/O ratio of 1.0. Spectral features for other cases are explained in the text. Spectra are shown at a resolution of $\lambda/\Delta\lambda = 600$.}
\label{fig:F8}
\end{figure*}

Atmospheres of increasingly shorter orbital periods are expected to be affected by stellar wind erosion \citep{Owen_2013,Lopez_2014,Lopez_2017,Fulton_2017}. For less massive worlds, such as ultra-short-period super-Earths, this likely results in extreme loss of volatile species, and even extensive tails of escaping particles (including silicates) \citep{Mura_2011,Castan_2011,Leger_2011}.  If the internal reservoir is unable to counteract depletion, the atmosphere will increase  its mean molecular weight. With the depletion of hydrogen, a \ce{CO}, \ce{CO2}, \ce{N2}, or \ce{SO2} atmosphere is  not an unusual outcome. In Figure \ref{fig:F8}, we investigate models that are severely depleted of hydrogen. As before, planets at two different orbital distances are shown with the oxygen abundance being determined via outgassing. In addition to hydrogen depletion, by varying the carbon mixing ratio, we explore the added impact of the C/O parameter.

For the close orbit cases the major difference caused by the depletion of hydrogen is the lack of \ce{H2O} in the atmosphere.  For a C/O ratio of 0.2 (grey TP profile and spectrum of Fig. \ref{fig:F8}), the thermal structure is similar to the cases discussed in Section \ref{sec:Results_I}. However, with no \ce{H2O} the excess oxygen makes \ce{SiO} the dominant constituent, followed by \ce{CO} and \ce{O2}. The depletion of hydrogen is yet another pathway in which \ce{SiO}-dominant atmospheres are possible. Due to the lack of \ce{H}, the \ce{H-} continuum is exchanged for the opacities of \ce{CO} and \ce{CO2}, with a major band of \ce{CO2} appearing at 4.5 \textmu m (faint upper spectrum in the bottom panel). \ce{SiO} still remains the only absorber at 9 \textmu m. If the C/O ratio is increased to 1.0 (blue curves), the formation of \ce{CO} becomes even more efficient, leaving \ce{SiO} nearly two orders of magnitude behind in volume mixing ratio (highlighted abundance curves in the top right panel). From a theoretical perspective, for C/O ratios close to unity, \ce{CO} atmospheres are easy to attain. This means that the  observability of the 5 \textmu m \ce{SiO} feature may not be feasible in such atmospheres. For comparison, we show a pure outgassing-produced silicate atmosphere (cyan). Between the volatile and silicate cases the 9 \textmu m \ce{SiO} feature remains; it might not be present    at 5 \textmu m. Detecting a combination of carbon oxides and \ce{SiO} could indicate that the atmosphere has    a high C/O ratio, with a significant outgassing component. 



\begin{figure}
    \centering
    \includegraphics[width=.49\textwidth]{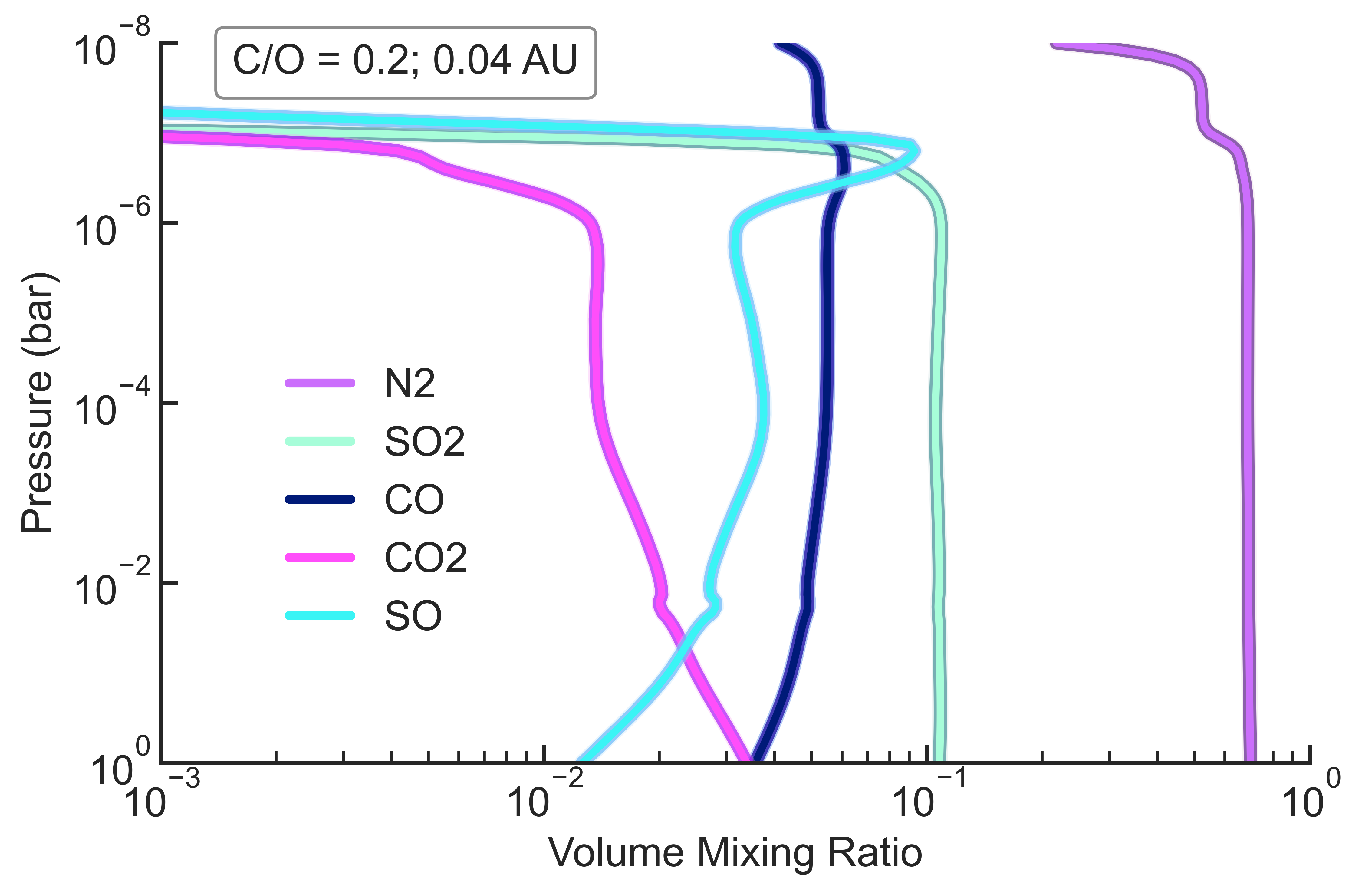}
    \caption{Five most abundant molecules in an atmosphere corresponding to a hydrogen-depleted case with C/O = 0.2 at 0.04 AU. They are, in decreasing abundance, \ce{N2}, \ce{SO2}, \ce{CO}, \ce{CO2}, and \ce{SO}. The relevant spectrum for this case is shown in Fig. \ref{fig:F8} (pink curve).}
\label{fig:F8Chem}
\end{figure}

In the less irradiated cases, with C/O = 0.2, the temperature profile is only inverted in the  uppermost region, similar to what was found in the original models (Fig. \ref{fig:F3}). However, the chemistry here is vastly different (see Figure \ref{fig:F8Chem}). The lack of outgassed oxygen allows for \ce{N2} to take over as the dominant species. Its abundance is also closely followed by sulphur, notably \ce{SO2}. While \ce{N2} is a weak absorber, the opacity of \ce{SO2} is significant, peaking at 4 and 7-9 \textmu m (grey lower spectrum of Fig.  \ref{fig:F8}, features are not marked). Sulphur-rich hot Venus-like atmospheres may be possible on irradiated rocky worlds, and should be taken into consideration for observations with MIRI LRS \citep{Schaefer_2011,Kane_2014,Zolotov,Liggings_2022}. Other slightly diminished yet still visible spectral features include those of \ce{CO} and \ce{CO2}. If the C/O ratio is increased to 1 (pink curves), \ce{N2} still remains as the dominant component; however, previously seen oxygen-rich species, such as \ce{SO2} are no longer abundant. Instead, \ce{PS} rises as the second most abundant molecule, causing an increase in shortwave opacity. The photosphere becomes severely inverted, resulting in large emission features. Aside from its shortwave component, \ce{PS} has potentially observable IR bands at 7 and 15 \textmu m (see Fig. \ref{fig:F2Extra} for its full opacity). Because of the strong \ce{VO} opacity at 10 \textmu m, this species may be interesting for observers.

\subsection{Observability of currently known targets}

Despite numerous observations with low- and high-resolution instruments, no definite detections of molecules have been made on any lava world. Thanks to  its broad spectral coverage and high sensitivity, JWST is expected to shed new light on the subject. So far, the performance of the telescope has surpassed all expectations \citep{Rigby_2022,JWST_CO_2022}. If short-period rocky planets do possess atmospheres, it is likely that JWST will be able to detect them   with greater confidence than anything before. Targets such as 55 Cnc e or K2-141 b are excellent labs to test silicate outgassing, and have been chosen to be observed during JWST's first cycle. If the observations are successful, many more targets are likely to follow. 

\begin{figure}
    \centering
    \includegraphics[width=0.49\textwidth]{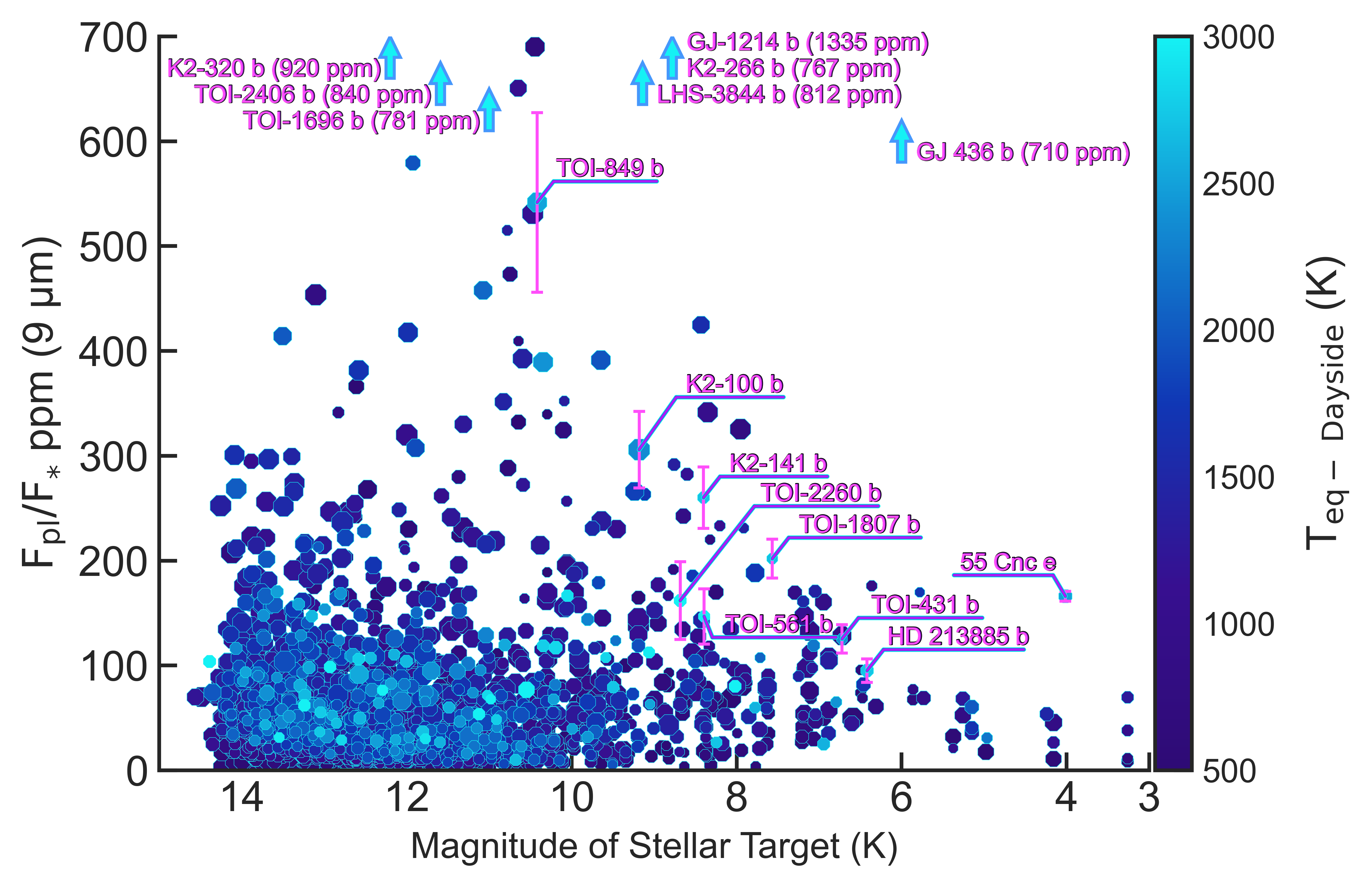}
    \caption{Currently confirmed super-Earths and sub-Neptunes plotted as a function of stellar magnitude (K) and emission flux of the planet at 9 \textmu m. The flux ratio here represents baseline emission modelled with dayside redistribution (f = 0.667), which is not affected by   absorbing species. 9 \textmu m is the wavelength where the largest \ce{SiO} feature is expected to manifest. A selection of favourable targets show \texttt{PANDEXO} simulated noise for the MIRI LRS instrument with 20 hours of observation time, binned to R=10.}
\label{fig:F11}
\end{figure}

In Figure \ref{fig:F11} we present currently confirmed planets, with a radius of 1-4 R${_\oplus}$, as a function of stellar magnitude and emission flux at 9 \textmu m. We note that here the flux ratio represents the baseline emission, not affected by possible absorption of present species. The simulated error bars are for 20 hours of observing time with MIRI LRS binned to a resolution of R = 10. The observability of a target will depend drastically not only on the contrast between the star and the planet, but also on the brightness of the system. As the surface temperature defines outgassed silicate abundances, the equilibrium temperature of the planet is an additional factor that needs to be considered. For less tenuous atmospheres, this condition may be somewhat relaxed as insulation of heat can allow for global magma oceans to be sustained at greater temperatures. Despite the ample number of discovered worlds, most orbit stars that are too dim to be good targets for atmospheric characterisation with the  JWST MIRI instrument. While the brightest systems, like 55 Cnc, have simulated noise of just a few ppm, at a stellar magnitude (K) of 9, it increases close to 50 ppm. Considering that such atmospheres only reach a few hundred ppm in the 9 \textmu m range, characterisation of any  features may be extremely difficult. However, there are a number of planets that are potentially good follow-up candidates for short-duration programmes.

One of the brightest and most studied super-Earths is 55 Cnc e, which will be observed with the  NIRCam and MIRI LRS instruments during JWST's first cycle \citep{HU_2021,Brandeker_2021}. Whether this planet possesses an atmosphere is currently debated, with a general consensus leading to an existing high-mean-molecular-weight atmosphere, possibly with an outgassed silicate component. Compositions dominated by \ce{CO}, \ce{N2}, or \ce{O2} are all possible;  low-metallicity \ce{H2}-rich atmospheres are less probable \citep{Demory_2016b,Angelo_2017,Zilinskas_2020,Zilinskas_2021},  though there have been claims of detected \ce{HCN}, which would indicate abundant \ce{H2} and a high C/O ratio \citep{Tsiaras_2016}. A recent reanalysis of Spitzer phase curve data of 55 Cnc e also suggests a high average dayside temperature of T${_{\star}}$= 3771 K, which may be caused by the presence of \ce{SiO} \citep{Mercier_2022}. Still, without observations of broad spectral coverage, conclusions for this planet cannot yet be drawn.

For high equilibrium temperatures several other targets  of various radii are of notable interest. Smaller rocky worlds, such as K2-141 b, TOI-1807 b, and TOI-561 b are ideal for observing silicates \citep{Hedges_2021,Dang_2021,Zieba_2022}. TOI-561 b is reported to have unusually low density, which could imply a large volatile component \citep{Lacedelli_2022}. Larger worlds include K2-100 b, TOI-849 b, and TOI-2260 b, all of which may also host volatile \ce{H2}-rich atmospheres with underlying magma oceans. Indicated via arrows are some of the highest contrast planets, including GJ-1214 b, GJ 436 b, LHS-3844 b, and several others. While the equilibrium temperature of these planets is generally too low to host magma oceans, an insulating atmosphere could force higher surface temperatures, and thus visible contamination of silicates.

The search for outgassed silicates is certainly not hindered by the lack of suitable targets, but rather by how small the spectral features are expected to be. While we do not model full spectra for any of the suggested targets, the emission features for most of them are     expected to closely mimic our presented 2 R${_\oplus}$ test cases. Generally, planets will not be observed for more than a few orbits, making spectral noise a considerable issue. With the  JWST MIRI LRS it should be possible to characterise the presence of \ce{SiO} on short-period rocky worlds and sub-Neptunes, but to do so will certainly be a grand challenge.

\section{Discussion}
\label{sec:Discussion}
\subsection{Importance of heat redistribution and stellar type}

\begin{figure}
    \centering
    \includegraphics[width=0.49\textwidth]{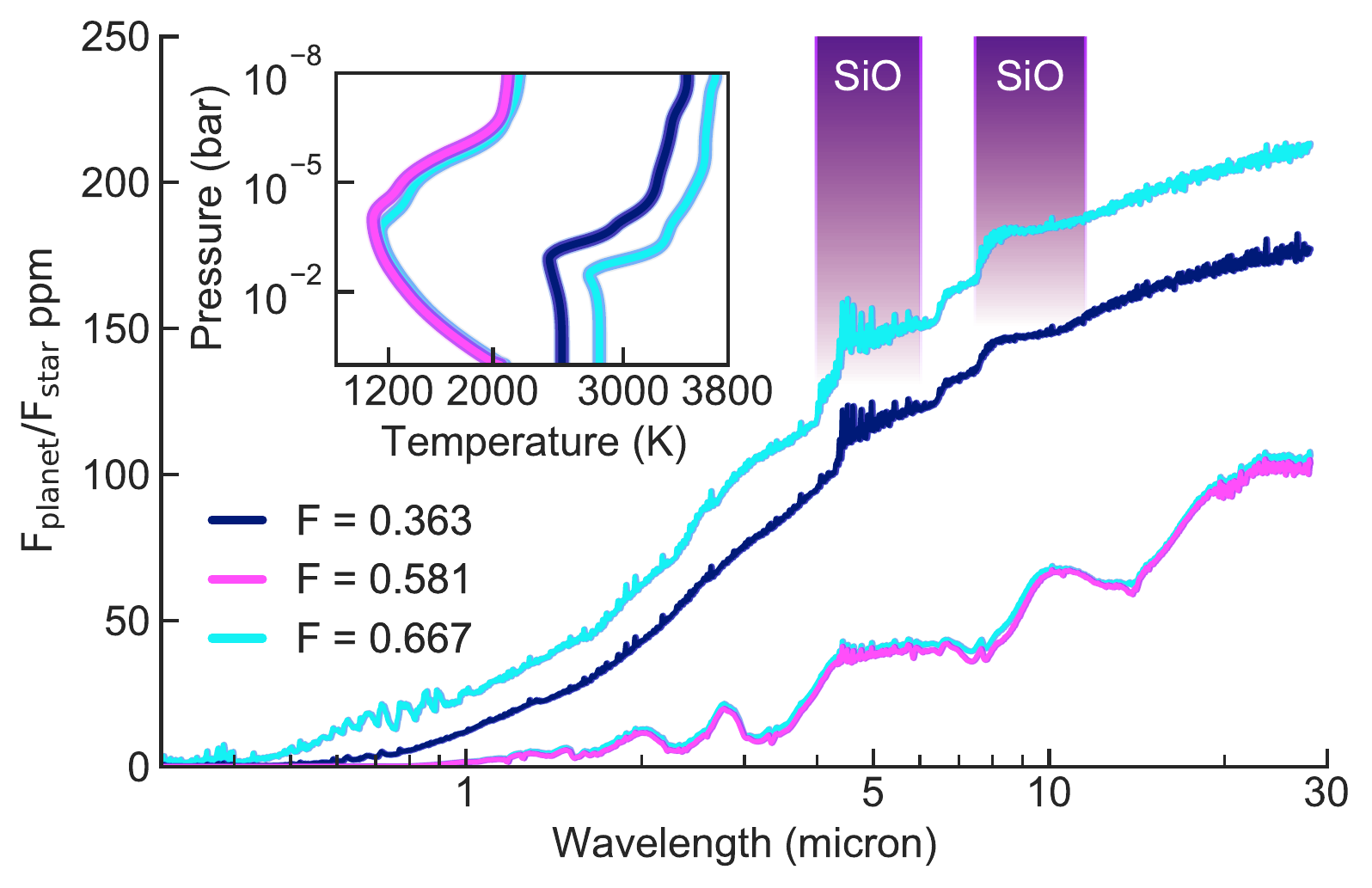}
    \caption{Temperature profiles and emission spectra of models with computed heat redistribution factor f at orbital distances of 0.015 and 0.04 AU. The atmospheres with f = 0.667 (both cyan curves) are arbitrarily set to represent the  dayside-confined redistribution of irradiation. Factors of f = 0.363 and f = 0.581 are calculated using the formulation of heat redistribution for rocky planets in \citet{Koll_2022}. The parameters determining heat transport efficiency are planetary equilibrium temperature, atmospheric surface pressure, and longwave optical depth. Spectra are shown at a resolution of $\lambda/\Delta\lambda = 600$.}
\label{fig:F9}
\end{figure}

The characterisation of lava worlds depends as much on the structure of the atmosphere as it does on the chemistry. 
While the spectral energy distribution of the host star largely determines the features of the temperature profile and the emission spectrum of the planet, the atmosphere itself is prone to physical processes that impact observability. Specifically, for 1D models, the efficiency of heat redistribution determines the total given energy budget, which subsequently decides much of the occurring chemistry. Approximations of this effect are made through a single factor f. Tenuous silicate atmospheres are expected to be inefficient in transporting heat, confining it to the dayside of the planet (f = 0.667--1.0). More volatile optically thick atmospheres are more efficient, where a significant fraction of the incoming stellar radiation can be deposited on the night side. A maximum full redistribution results in f = 0.25. Using the analytical heat redistribution scaling theory from \citet{Koll_2022}, in Figure \ref{fig:F9} we show the possible effect that this may have on the observability of silicates. 

In general, we assume that our modelled planets do not redistribute heat efficiently. Most of the models are set to use f = 0.667; however, using a scaling theory derived for tidally locked worlds, we find that strongly irradiated atmospheres with volatiles can become very efficient at transporting heat in cases reaching as high as f = 0.363. While not attaining global redistribution (f = 0.25), this severely impacts the total energy budget, reducing the surface temperature and thus silicate observability. It is no surprise that volatile atmospheres increase the efficiency, but the large magnitude of f for just 1 bar of an atmosphere does indicate that even a small amount of volatiles can have a severe impact. On the contrary, planets at larger orbital distances do not show a large increase in redistribution efficiency; they  keep most of the energy confined to the dayside. Since the primary mechanism of heat transport is the atmosphere, phase curve measurements can indicate its significance. Detecting temperature offsets or high nightside flux could indicate that the planet has retained a substantial volatile-rich atmosphere. The unusual super-Earth 55 Cnc e has been found to show signs of this \citep{Demory_2016b}.

\begin{figure}
    \centering
    \includegraphics[width=0.49\textwidth]{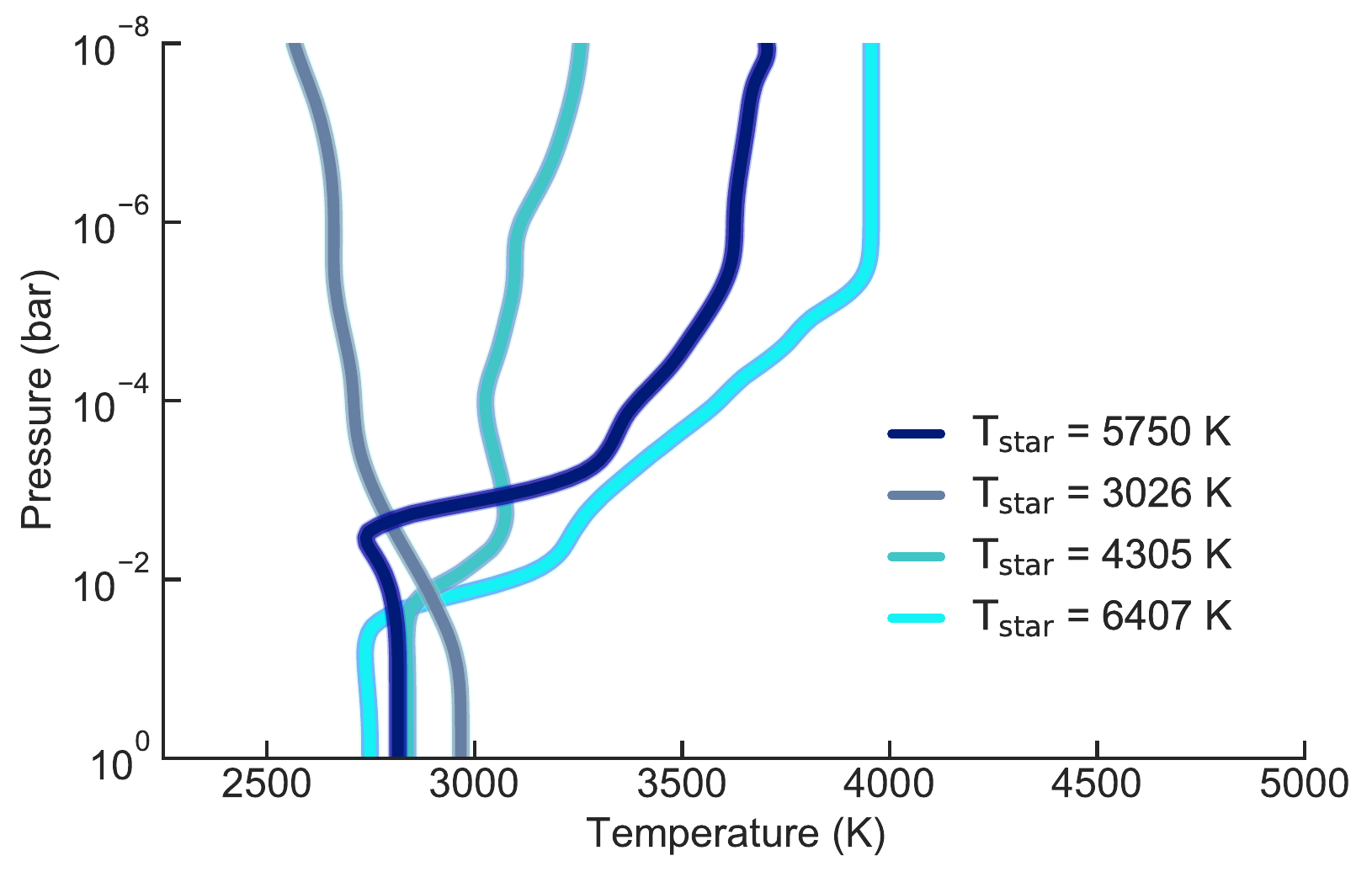}
    \caption{Temperature profiles computed consistently with atmospheric chemistry for different stellar-type hosts. In each case the planet is placed at an equivalent irradiation distance from the star. T${_{\star}}$= 3026 K spectrum represents GJ 1214 and is   from the MUSCLES database. The other three are modelled using PHOENIX spectra of the denoted temperature.}
\label{fig:F10}
\end{figure}

Since emission of the planet probes its thermal profile, observability is also greatly impacted by the spectrum of the host star. In Figure \ref{fig:F10}, we take one of our solar cases and compare it to models of planets placed around stars of different types, but at an equivalent irradiation distance. With increasing stellar temperature, its spectrum is shifted towards shorter wavelengths, causing greater incident UV flux. Going from a G-type to a typical K-type star (T$_{\star}$= 4305 K) the inversions weaken drastically. Atmospheres around cool M dwarfs are not likely to contain  a deep inversion that strongly affects the emitting photosphere. For the characterisation of lava worlds through emission spectroscopy this may present a slight problem since strong inversions are one of the key characteristics of silicate-rich atmospheres that may help to identify them.

\section{Conclusion}
\label{sec:Conclusions}

In preparation for the JWST observations, we have used consistent outgassing equilibrium chemistry and radiative-transfer models to assess the possibility of detecting silicates in volatile atmospheres of super-Earths and smaller sub-Neptunes. Placing a hypothetical 2 R${_\oplus}$ planet at varying orbital distances around a Sun-like star, we have explored a wide variety of viable atmospheric compositions, rich in \ce{H}, \ce{C,} and \ce{N}, that also include an outgassed silicate component, for example \ce{Si}, \ce{O}, or  \ce{Ti}. We modelled atmospheres of up to 10 bar surface pressure, varied in metallicity and C/O ratio. A major assumption made in this work is that the atmosphere is in complete equilibrium with the underlying melt. Our results are intended to guide observers towards potentially detectable species that would help characterise worlds with exposed or concealed lava oceans. Our key findings are as follows:

   \begin{enumerate}

    \item For emission spectroscopy with JWST, \ce{SiO} is likely to remain the only characterisable species that could indicate strong outgassing from an underlying magma ocean. However, in atmospheres containing volatiles, the main opacity bands of \ce{SiO} at 5 and 9 \textmu m can be heavily suppressed. Only the most irradiated worlds with melt temperatures > 2500--2800 K are expected to show super-solar abundances of silicates (assuming 1 bar of volatiles). Contrary to what has been found for pure lava worlds, we do not find features of \ce{SiO2} to be of significance. Very high temperature melts may also result in broad shortwave features, arising mainly from \ce{TiO} and several other outgassed species. Ultimately, the visibility of silicates in volatile atmospheres   largely depends on the atmospheric properties and the efficiency of its interaction with the melt.
    \\
    
    \item Thermal inversions are prominent in atmospheres contaminated with silicates. We find that numerous outgassed silicates cause deep inversions that affect the photosphere, even if the atmosphere has strong infrared absorption originating from volatiles. The largest contributors to shortwave opacity are \ce{TiO}, \ce{SiO}, \ce{MgO}, \ce{Fe,} and \ce{Mg}. We also find that alkali metals, metal hydrates, and, in certain cases, \ce{PS} or \ce{VO} can become a source of inversions. Because outgassing scales exponentially with the temperature of the melt, planets without insulating atmospheres and at larger orbital distances are not likely to show strong inversions  due to silicates.
    \\
    \item The presence of hydrogen can affect the observability of silicates, including  \ce{SiO}. Our models show that added hydrogen results in the formation of a large quantity of  hydrocarbons and \ce{H2O}. Chemically, \ce{H2O} and \ce{SiO} are direct competitors for the outgassed oxygen; however, \ce{SiO} is much less prone to thermal dissociation, making it more prominent in the lower-pressure inversion-affected regions. On strongly irradiated worlds, the continuum of \ce{H-} can also heavily obscure the 5 and 9 \textmu m \ce{SiO} features. 
    \\
    \item The C/O ratio has a large effect on \ce{SiO}. Even in \ce{H2}-rich atmospheres, the formation of \ce{CO} due to a C/O ratio $\gtrsim$ 1.0 can result in a drastic reduction of silicate oxides. Chemically, \ce{CO} takes priority for oxygen, affecting all other oxides. This can consequently result in \ce{Si} bonding with \ce{H} instead of \ce{O}, forming \ce{SiH}, potentially making it a species of interest for observations. Detecting \ce{CO} at 4.5 \textmu m and \ce{SiO} at 9 \textmu m could indicate an atmosphere with no hydrogen and a high C/O ratio. 
    \\
    \item Atmospheric pressure, insulation, and redistribution of heat are major factors in deciding whether volatile atmospheres are contaminated with silicates. Our models show that atmospheres of 10 bar with internal temperature enabled become convective, resulting in a large increase in surface temperature. Exponential scaling of outgassing can lead to \ce{SiO} becoming a dominant species, even in substantial volatile atmospheres. On the contrary, effective heat redistribution can reduce surface temperatures. Using a scaling theory for tidally locked planets, we find that even small volatile atmospheres are efficient at transporting heat, and thus lowering surface temperatures. Observations of silicates can therefore provide important constraints on the underlying melt properties and the balance between insulation and redistribution of heat.

   \end{enumerate}

Shrouded magma oceans are expected to severely contaminate even substantial volatile envelopes. Detection of silicon oxides would be a major step in allowing us to put strong constraints on the efficiency of the interaction between the magma and the overlying atmosphere. We thus strongly recommend to observe irradiated super-Earths and sub-Neptunes for the presence of \ce{SiO} and \ce{SiO2}.

\begin{acknowledgements}
 
\\
We acknowledge funding from the European Research Council under the European Union’s Horizon 2020 research and innovation program under grant agreement No 694513. We thank Matej Malik for the insightful discussion on \texttt{HELIOS}. We are also grateful for the comments of the editor and the anonymous referee, which helped improve the quality of this paper.
\\
\\
Software used in this work:
\texttt{HELIOS-K} \citep{Grimm_2015,Grimm_2021}; \texttt{HELIOS}\citep{Malik_2017,Malik_2019}; \texttt{FASTCHEM} \citep{Stock_2018}; \texttt{LavAtmos} \citep{Buchem_2022}; \texttt{petitRADTRANS} \citep{Molliere_2019,Molliere_2020}; \texttt{PANDEXO} \citet{Batalha_2017}; \texttt{numpy} \citep{Harris_2020}; \texttt{matplotlib} \citep{Hunter_2007}; \texttt{seaborn} \citep{Waskom_2021}; \texttt{astropy} \citep{Astropy_2022}.
\\
\\
Supplementary material is available on request from the author.

\end{acknowledgements}

\bibliographystyle{aa}
\bibliography{references}
\onecolumn
\clearpage
\begin{appendix} 
\section{Opacity Data}
\label{appendixA}

\begin{longtable}{lllll} 
 \caption[]{Description of the opacities used to calculate temperature profiles and emission spectra.}
 \\
 \label{table:opacities}
Species & Source & Line list & Line List Reference
    \\
  \hline
  \hline
  \\
  \ce{Al} & DACE$^a$ & VALD & \citet{Ryab_2015}\\
  \ce{AlH} & HELIOS-K & AlHambra & \citet{Yurchenko_2018b}\\
  \ce{AlO} & HELIOS-K$^b$ & ATP & \citet{Patrascu_2015} \\
  \ce{Ca} & DACE & VALD & \citet{Ryab_2015}\\
  \ce{CaH} & HELIOS-K & MoLLIST & \citet{Li_2012,Bernath_2020}\\
  \ce{CaO} & HELIOS-K & VBATHY & \citet{Yurchenko_2016}\\
  \ce{Fe} & DACE & VALD & \citet{Ryab_2015}\\
  \ce{K} & DACE & VALD & \citet{Ryab_2015}\\
  \ce{Mg} & DACE & Kurucz & \citet{Kurucz_1992}\\
  \ce{MgH} & HELIOS-K & MoLLIST & \citet{GharibNezhad_2013,Bernath_2020}\\
  \ce{MgO} & HELIOS-K & LiTY & \citet{Li_2019}\\
  \ce{Na} & DACE & VALD & \citet{Ryab_2015}\\
  \ce{NaH} & HELIOS-K & Rivlin & \citet{Rivlin_2015}\\
  \ce{Si} & DACE & VALD & \citet{Ryab_2015}\\
  \ce{SiH} & HELIOS-K & Sightly & \citet{Yurchenko_2018a}\\  
  \ce{SiO} & HELIOS-K & SiOUVenIR & \citet{Yurchenko_2022}\\
  \ce{SiO2} & DACE & OYT3 & \citet{Owens_2020}\\
  \ce{Ti} & DACE & VALD & \citet{Ryab_2015}\\
  \ce{TiH} & HELIOS-K & MoLLIST & \citet{Burrows_2005,Bernath_2020}\\
  \ce{TiO} & HELIOS-K & Toto & \citet{McKemmish_2019}\\
    \\
     Volatiles \\
  \hline    
  \\
  \ce{CO} & DACE & Li2015 & \citet{Gang_2015}\\
  \ce{CO2} & DACE & HITEMP \& UCL-4000$^c$ & \citet{Rothman_2010,Yurchenko_2020}\\
  \ce{CH3} & DACE & AYYJ & \citet{Adam_2019}\\
  \ce{CH4} & DACE & YT34to10 & \citet{Yurchenko_2017}\\
  \ce{C2H2} & DACE & aCeTY & \citet{Chubb_2020}\\
  \ce{C2H4} & DACE & MaYTY & \citet{Mant_2018}\\  
  \ce{CN} & HELIOS-K & Trihybrid & \citet{Syme_2021}\\  
  \ce{H2O} & DACE & POKAZATEL & \citet{Polyansky_2018}\\  
  \ce{HCN} & HELIOS-K & Harris & \citet{Barber_2014}\\  
  \ce{HS} & HELIOS-K & GYT & \citet{Gorman_2019}\\  
  \ce{H2S} & DACE & AYT2 & \citet{Azzam_2016}\\  
  \ce{NH3} & DACE & CoYuTe & \citet{Coles_2019}\\ 
  \ce{OH} & DACE & HITEMP & \citet{Rothman_2010}\\  
  \ce{S} & DACE & VALD & \citet{Ryab_2015}\\ 
  \ce{CS} & ExoMol$^d$ & JnK & \citet{Paulose_2015}\\ 
  \ce{NS} & ExoMol & SNaSH & \citet{Yurchenko_2018c}\\ 
  \ce{SO2} & ExoMol & ExoAmes & \citet{Underwood_2016a}\\ 
  \ce{SO3} & ExoMol & UYT2 & \citet{Underwood_2016b}\\ 
  \ce{PH3} & DACE & SAlTY & \citet{Sousa_2015}\\ 
  \ce{PS} & HELIOS-K & POPS & \citet{Prajapat_2017}\\ 
  \ce{VO} & HELIOS-K & VOMYT & \citet{McKemmish_2016}\\ 
\\
    Scattering and Continuum \\
  \hline    
  \\
  \ce{H}, \ce{H2}, \ce{H2O}, \ce{He}, \ce{O2} & Scattering & \\  
  \ce{H-} & Continuum (bf \& ff) &  & \citet{John_1988,Gray_2008}\\  
  \ce{H2-H2} & HELIOS-K & HITRAN & \citet{Gordon_2017}\\ 
  \ce{H2-He}, \ce{O2-O2} & petitRADTRANS & & \citet{Molliere_2019,Molliere_2020}\\   
  \\
  \hline
  \multicolumn{4}{p{17.4cm}}{\footnotesize$^{a}$ DACE database https://dace.unige.ch/; \footnotesize$^{b}$ Opacities are computed with \texttt{HELIOS-K} https://github.com/exoclime/HELIOS-K \citep{Grimm_2015,Grimm_2021};  \footnotesize$^{c}$ We use HITEMP2010 for temperature profiles and UCL-4000 for emission spectra; \footnotesize$^{d}$ For ExoMol opacities we make use of the conversion work done by \citet{Chubb_2021}, which are only used for generating emission spectra. In general, we make heavy use of the DACE \citep{Grimm_2015,Grimm_2021}, ExoMol \citep{Chubb_2021}, Kurucz \citep{Kurucz_1992}, VALD3 \citep{Ryab_2015} and HITRAN \citep{Gordon_2017} databases.}
\\

\end{longtable}
\twocolumn

\end{appendix}
\end{document}